\documentclass[journal]{IEEEtran}
\usepackage{booktabs}
\usepackage{amsfonts}
\usepackage{amsmath}
\usepackage{graphicx}
\usepackage{multirow}
\usepackage{algorithm}
\usepackage{diagbox}
\usepackage{yhmath}
\usepackage{algpseudocode}
\usepackage{subfig}
\usepackage{url}
\usepackage{enumitem}
\ifCLASSINFOpdf
\else
\fi
\begin{document}
%
\title{Data Augmentation for Enhancing EEG-based Emotion Recognition with Deep Generative Models}
%
%
%

%
%

\author{Yun~Luo,
        Li-Zhen~Zhu,
        Zi-Yu~Wan,
        and~Bao-Liang~Lu*,~\IEEEmembership{Senior~Member,~IEEE}
\thanks{Y. Luo, L.-Z. Zhu, Z.-Y. Wan, and B.-L. Lu are with the Center for Brain-Like Computing and Machine Intelligence, Department of Computer Science and Engineering, Shanghai Jiao Tong University, Shanghai 200240, China, also with the Key Laboratory of Shanghai Education Commission for Intelligent Interaction and Cognitive Engineering, Shanghai Jiao Tong University, Shanghai 200240, China, and with the Brain Science and Technology Research Center, Shanghai Jiao Tong University, Shanghai 200240, China. * Corresponding author: Bao-Liang Lu (bllu@sjtu.edu.cn).  }}

\markboth{Journal of \LaTeX\ Class Files,~Vol.~14, No.~8, August~2015}%
{Shell \MakeLowercase{\textit{et al.}}: Bare Demo of IEEEtran.cls for IEEE Journals}
%



\maketitle

\begin{abstract}
The data scarcity problem in emotion recognition from electroencephalography (EEG) leads to difficulty in building an affective model with high accuracy using machine learning algorithms or deep neural networks. Inspired by emerging deep generative models, we propose three methods for augmenting EEG training data to enhance the performance of emotion recognition models. Our proposed methods are based on two deep generative models, variational autoencoder (VAE) and generative adversarial network (GAN), and two data augmentation strategies. For the full usage strategy, all of the generated data are augmented to the training dataset without judging the quality of the generated data, while for partial usage, only high-quality data are selected and appended to the training dataset. These three methods are called conditional Wasserstein GAN (cWGAN), selective VAE (sVAE), and selective WGAN (sWGAN). To evaluate the effectiveness of these methods, we perform a systematic experimental study on two public EEG datasets for emotion recognition, namely, SEED and DEAP. We first generate realistic-like EEG training data in two forms: power spectral density and differential entropy. Then, we augment the original training datasets with a different number of generated realistic-like EEG data. Finally, we train support vector machines and deep neural networks with shortcut layers to build affective models using the original and augmented training datasets. The experimental results demonstrate that the augmented training datasets produced by our methods enhance the performance of EEG-based emotion recognition models and outperform the existing data augmentation methods such as conditional VAE, Gaussian noise, and rotational data augmentation.
\end{abstract}

\begin{IEEEkeywords}
Emotion recognition, EEG, GAN, VAE, deep generative model, data augmentation.
\end{IEEEkeywords}

%
\IEEEpeerreviewmaketitle

\section{Introduction}
Emotion plays a significant role in how people think, behave, and communicate.
Artificial emotional intelligence, which is also known as emotion AI or affective computing, focuses on developing devices and systems that can automatically recognize human emotion and has attracted considerable attention very recently \cite{emotionAI1,emotionAI2}. For example, integrating emotion assessment in human-computer interaction systems with emotion recognition can make machines more intelligent and provide more humanized interactions. Moreover, studies have shown that some mental diseases, such as depression and autism, are relevant to emotions \cite{Bocharov2017Depression}. The introduction of emotion AI to these studies can create a high potential for treating psychiatric diseases. Because emotion AI has many potential applications, attention is being focused on the possibility of recognizing emotions from different behavioral cues, such as facial expression \cite{Liu2017A}, posture \cite{Garber2012Using}, voice \cite{Tanja2009Emotion}, and neurophysiological signals \cite{Samara2017Feature}. Among these signals, electroencephalography (EEG) has been demonstrated as one of the most reliable signals due to its high accuracy and objective benefits. In recent years, EEG-based emotion recognition has attracted widespread attention from academics and industries \cite{Wang2014Emotional,McFarland_2016,Alarc1949Emotions,Craik_2019,9043472}. Researchers have made considerable progress in feature extraction and model construction. However, these studies are faced with a problem: the lack of training data.

Compared with visual and audio signals, which can be easily accessed from standard datasets, data acquisition is still one of the bottlenecks in EEG-based emotion recognition tasks. There are mainly five reasons: a) The price of EEG acquisition devices for research is quite high. Additionally, EEG-based emotion recognition experiments are time-consuming and require tedious preparations, such as skin cleaning and gel injection, which makes it difficult to conduct many experiments. b) These experiments cannot last for a long time because the subjects may feel uncomfortable wearing EEG acquisition devices. Therefore, it is difficult to acquire large-scale labeled EEG data in one experiment. c) The raw EEG data are usually mixed with noise and various artifacts, and researchers have to discard some bad channels and data, which aggravates the data scarcity problem. d) It is difficult to collect precisely labeled data since the subjects may not evoke emotion well in emotion recognition experiments. e) There are only a few public EEG-based emotion recognition datasets, such as SEED\footnote{http://bcmi.sjtu.edu.cn/$\sim$seed/index.html} \cite{zheng2015investigating}, DEAP\footnote{http://www.eecs.qmul.ac.uk/mmv/datasets/deap/} \cite{koelstra2012deap}, DREAMER \cite{7887697}, MAHNOB-HCI\footnote{https://mahnob-db.eu/hci-tagging/} \cite{5975141}, and MPED \cite{8606087}. Moreover, the scales of these datasets are much smaller than those of public image datasets (e.g., ImageNet). These factors limit the quantity of labeled training data for EEG-based emotion recognition and hinder the performance of emotion recognition models trained by machine learning algorithms and deep neural networks.

It is common sense that a machine learning model will be more accurate when it can access more training data. For example, the release of the trillion-word corpus by Google improves text-based models \cite{UnreasonableData}. Machine learning models can be more robust and reliable when learning more effective features from sufficient training data, especially for deep learning models that need a vast quantity of training data. Deep learning models have recently achieved remarkable results in the fields of computer vision, speech recognition, and natural language processing due to the accessibility of large datasets \cite{Zhang2018A}.

In the field of EEG-based emotion recognition, Zheng and Lu used deep neural networks to recognize three emotions and reached a compelling accuracy \cite{zheng2015investigating}. In their work, they only applied a two-layer deep belief network. The achievements in the image, speech, and natural language processing fields indicate that there is considerable room for further studying the problem of EEG-based emotion recognition by leveraging the ability of deeper neural networks. However, compared with shallow layer models, deep-layer models use more parameters and require a large number of labeled training data to explore the potentials of deep neural networks. Consequently, the primary issue that should be addressed in EEG-based emotion recognition is to acquire sufficient and high-quality training data.

Generating artificial data by applying a transformation from the original data is one of the conventional solutions to solving the data scarcity problem. This approach is called data augmentation. Recently, various data augmentation methods have been applied to generate EEG data \cite{Krell2017Rotational,Lotte2015Signal,Fang2018Data}. Some researchers have generated EEG data by applying a geometric transformation to the original data and reported the performance of classifiers improved by adding the generated data. Other researchers have focused on using deep generative models to generate artificial EEG data \cite{Hartmann2018EEG}. Compared with signal-level transformation through geometric transformation, the deep generative model learns the representation of the real distribution at a deeper level, which leads to better classification results. However, the performance of the classifier after data augmentation were not demonstrated \cite{Hartmann2018EEG}. In our previous study, we generated realistic-like EEG features by taking advantage of GANs \cite{luo2018embc}. Then, we compared the performance of affective models without and with appending the generated data to the training dataset. The experimental results demonstrated that the GAN-based data augmentation method could improve the performance of affective models. In this paper, we further explore the generative methods based on the above achievement.

It is difficult to collect completely pure EEG signals due to the low signal-to-noise ratio (SNR). In addition, it is common for classifiers to handle the high-level features of EEG data in EEG-based emotion recognition tasks. Therefore, this work focuses on generating power spectral density (PSD) and differential entropy (DE) features, which are two commonly used features in emotion recognition tasks \cite{zheng2015investigating,duan2013differential,yang2018eeg-based}.

\begin{figure}[t]
\centering
\subfloat[All of the generated data are used to augment the training data set.]{\includegraphics[width=0.49\textwidth]{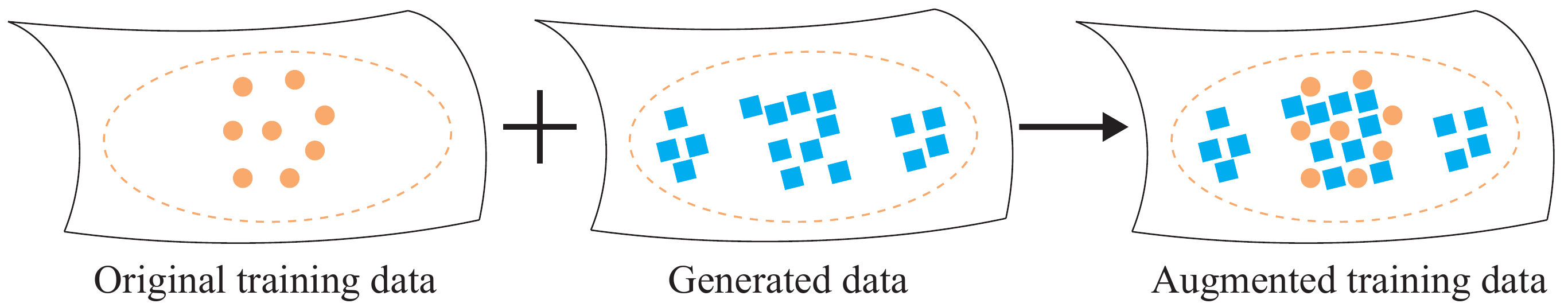}}\hfill
\subfloat[The generated data with high quality are selected to augment the training data set.]{\includegraphics[width=0.49\textwidth]{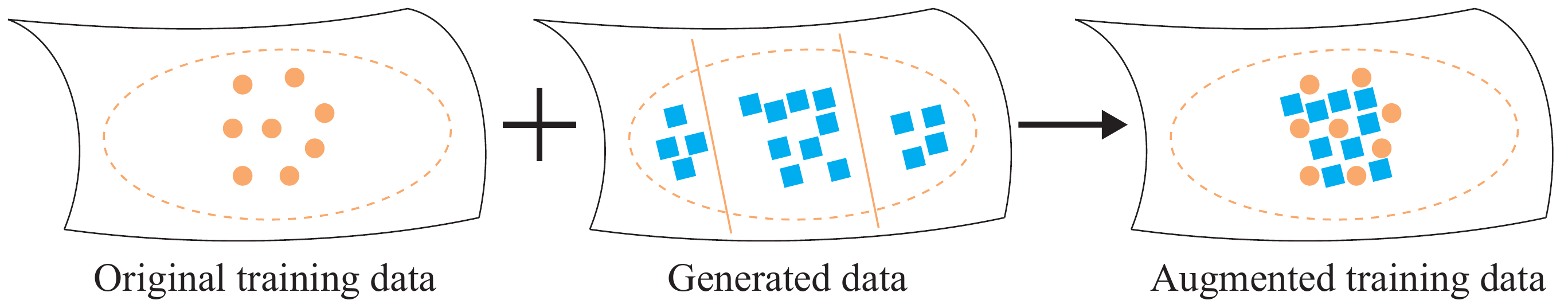}}\hfill
\caption{Illustration of two strategies of data augmentation used in this work. }
\label{fig:1}
\end{figure}

The work includes two emerging deep generative models: variational autoencoder (VAE) \cite{kingma2014auto-encoding} and Wasserstein generative adversarial network (WGAN) with gradient penalty \cite{arjovsky2017wasserstein,gulrajani2017improved}. We propose two data augmentation strategies: full usage of generated data and partial usage of generated data. As illustrated in Fig.1, the basic ideas behind the full usage strategy and partial usage strategy are to use all of the generated data and select part of the generated data. Since we cannot guarantee that all of the generated data have high qualities, it is important for us to decide how to use the generated data. For the full usage strategy, we propose conditional Wasserstein GAN (cWGAN) to control the category of the generated data. Then, we append all of the generated data to the original training dataset without considering their quality.

For the partial usage strategy, we propose two methods called selective VAE (sVAE) and selective WGAN (sWGAN) to generate data. In these two methods, the generated data are unlabeled. By applying SVMs as classifiers, we choose the generated data with high classification confidence and append the selected data to the original training dataset. Unlike images, the generated EEG features are high-dimensional data and are intractable for humans to judge the quality of the generated data. Therefore, these two methods are based on a simple idea, and the generated data are regarded as high quality when they are classified with high classification confidence by a classifier trained by the original dataset. We use two conventional pattern classifiers, SVMs and deep neural networks with shortcut layers, to train affective models on two public EEG datasets widely used for emotion recognition: SEED and DEAP.

For a comparison study, we introduce three conventional data augmentation methods for EEG-based emotion recognition: conditional VAE (cVAE) \cite{kingma2014auto-encoding}, which adopts a similar generated strategy as cWGAN, and Gaussian noise method (Gau), which augments the datasets by adding Gussain noise to the original data \cite{Fang2018Data}, and rotational data augmentation method (RDA), which generates new data by applying a geometric rotation to the original data \cite{Krell2017Rotational}. We perform a systematic experimental study to compare the proposed methods with these conventional methods. We use 5-fold cross-validation to measure the classification performance of different augmented methods. The proposed framework is illustrated in Fig. 2.

\begin{figure*}[ht]
\centering
\includegraphics[width=0.98\textwidth]{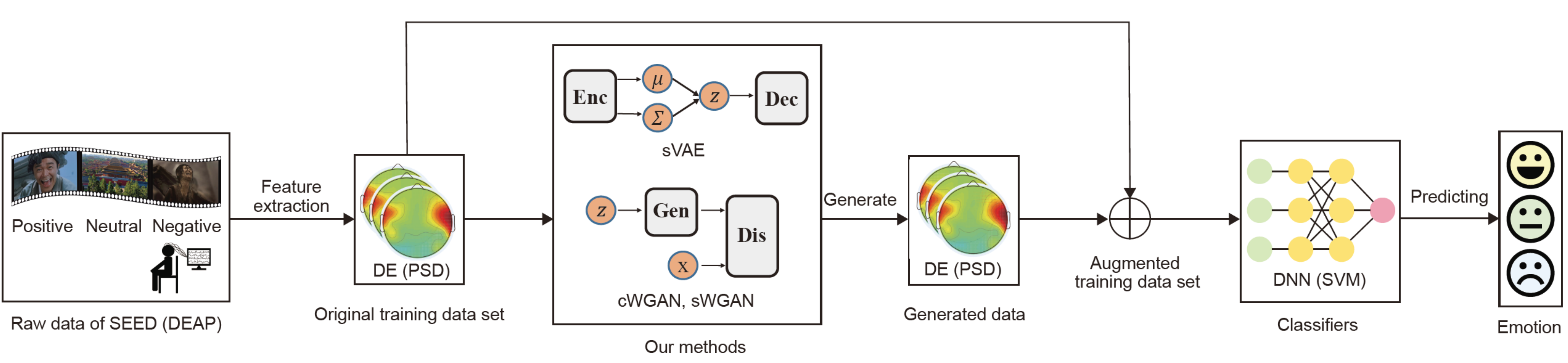}
\caption{Illustration of our proposed data augmentation framework. First, we extract the DE feature and PSD feature from two EEG-based emotion recognition datasets, SEED and DEAP, respectively. Second, we use our proposed three methods to generate realistic-like data and augment the original training dataset. Finally, we evaluate the performance of the proposed methods using SVMs and DNNs with shortcut layers.}
\label{fig:2}
\end{figure*}

The main contributions of this paper lie in the following aspects:
\begin{enumerate}[]
\item To the best of our knowledge, we adopt deep generative methods to augment EEG training data for emotion recognition for the first time.
\item We propose three methods for generating EEG data based on different generative methods and two different strategies for using the generated data.
\item We carry out a systematic comparison between different features, different generative methods and different classifiers on two EEG datasets. And the experimental results demonstrate that our proposed methods could make the affective models have better performance on EEG-based emotion recognition.
\end{enumerate}

The rest of the paper is organized as follows. Section 2 provides an overview of related work on generative methods, data augmentation methods for EEG-based emotion recognition, and a brief introduction to deep neural networks. In section 3, we introduce different methods in detail. Section 4 describes the two datasets, SEED and DEAP, and presents the details of our experimental settings. A systematic comparison between different methods and the efficiency of our proposed methods by conducting a series of data augmentation experiments is presented in section 5. Finally, in section 6, we present conclusions about our work.

\section{Related Work}
\label{sec:related work}
In this section, we briefly introduce relevant work on EEG-based emotion recognition, deep generative methods, data augmentation methods, and deep neural networks.
\subsection{EEG-based Emotion Recognition}
EEG-based emotion recognition has received considerable attention. M{\"u}hl \emph{et al.} introduced affective factors into traditional brain-computer interfaces (BCIs) \cite{Zander2012Context} and presented the definition of affective brain-computer interfaces (aBCIs) \cite{M2014A}. They summarized neurophysiology-based affect detection methods in recent years and discussed the limitations and challenges in this emerging field. Alarcao and Fonseca surveyed different EEG-based emotion recognition methods and compared the main aspects involved in the recognition process, including stimuli, feature extraction methods, and classifiers \cite{alarcao2017emotions}. Jenke \emph{et al.} reviewed feature extraction and selection methods for 33 EEG-based emotion recognition studies \cite{jenke2014feature}. Petrantonakis and Hadjileontiadis presented a novel EEG-based feature extraction technique by employing higher-order crossings analysis \cite{highOrder}.

Researchers used different emotion stimuli in their studies, such as music \cite{music1,music2}, image \cite{sohaib2013evaluating,petrantonakis2011a} and movie clip \cite{zheng2015investigating,koelstra2012deap}. Among these stimuli, the movie clip is believed to be one of the most efficient methods for eliciting human emotion. Koelstra \emph{et al.} developed a publicly available EEG-based emotion dataset called DEAP by recruiting 32 subjects to watch 40 music videos. Zheng and Lu required 15 subjects to watch 15 selected Chinese movie clips to elicit three emotions: happy, sad, and neutral (SEED dataset) \cite{zheng2015investigating}. Then, they performed a systematic comparison between various feature extraction, feature selection, feature smoothing and classification methods in a three-category EEG-based emotion recognition task and showed the stable patterns of EEG in this task \cite{zheng2017identifying}. In addition, they developed a multimodal framework for emotion recognition called EmotionMeter \cite{zheng2019emotionmeter}. In their work, they designed a six-electrode placement to collect EEG in an emotion recognition task including four emotions: happy, sad, neutral and fear. By combining EEG and eye movement signals, they achieved appealing recognition results \cite{zheng2019emotionmeter}. Zhao \emph{et al.} also adopted this framework and extended it to five-category emotion recognition: happy, sad, neutral, fear, and disgust \cite{zhao2018emotion}. Their results demonstrated the effectiveness of EEG signals for emotion recognition tasks.

\subsection{Deep Generative Methods}
Generative models aim to learn the data distribution of a given dataset using unsupervised learning to generate new data with some variations and have been widely studied in the field of machine learning. Recent advances in parameterizing these models using deep neural networks have allowed them to scale to diverse data, including images, text, and speech. Two of the most promising and efficient deep generative models are the variational autoencoder (VAE) \cite{kingma2014auto-encoding} and generative adversarial network (GAN) \cite{goodfellow2014generative}.

As a variation in the autoencoder, VAE aims at generating new data only based on the given data \cite{kingma2014auto-encoding}. It solves the variational inference problem that maximizes the marginalized data likelihood by using a generative network (encoder) and a recognition network (decoder). At the end of the training, the encoder can generate realistic-like data. VAEs have shown great potential in generating different data \cite{salimans2015markov,kulkarni2015deep,gregor2015draw}. Considering the generative ability of VAEs, we choose the vanilla VAE as one of our generative models.

As an emerging topic, GAN has attracted growing interest since it was first proposed by Goodfellow \emph{et al.} \cite{goodfellow2014generative}. The idea of GAN is to sample noise from distributions such as Gaussian and transform them into real data distributions. GANs are based on a mini-max game theory that aims to find the Nash equilibrium between the two components, generator and discriminator. The generator learns the data distribution, while the discriminator estimates the probability that a sample comes from the real data distribution or the noise distribution. During the game, the generator tries to fool the discriminator by generating realistic-like data, while the discriminator tries not to be fooled by improving the discriminating capability. After the adversarial process, the generator can produce high-quality faked data.

GANs have witnessed great success in recent years. Considering that the original GAN has no control over modes of the generated data, Mirza and Osindero added the label as an additional parameter to the generator and the discriminator to control the category of the generated data \cite{Mirza2014Conditional}. A similar idea was also adopted in InfoGAN by introducing latent codes \cite{infoGAN}. GANs also show promise in generating realistic-like data in specific fields. Ledig \emph{et al.} proposed SRGAN for image superresolution \cite{ledig2017photo-realistic}. By leveraging the ability of GANs, they could create high-resolution images from a single low-resolution image. Wu \emph{et al.} proposed 3D-GAN to generate 3D objects from a probabilistic space using volumetric convolutional networks and generative adversarial networks \cite{wu2016learning}. They focused on video generation by taking advantage of GANs and achieved considerable results. In addition, GANs have also been applied to the generation of dialogue \cite{Li2017Adversarial}, electronic health records (EHRs) \cite{Choi2017Generating}, and polyphonic music \cite{Mogren2016C}.

Although GANs have demonstrated great generation abilities, they have some problems, such as nonconvergence, mode collapse, and diminished gradient. Chief among them is training stability (nonconvergence), which is mainly caused by the adversarial game. Some pioneering works focus on fixing this problem. Radford \emph{et al.} reported some network architecture recommendations about GANs and designed a sophisticated network called DCGAN \cite{Radford2015Unsupervised}. Their work made a great contribution to solving the instability problem of GANs' training process. However, DCGAN was designed for image generation and requires specific design techniques, which limited it to scale to other fields. For this reason, other researchers have focused on altering the structure \cite{began} or the loss function \cite{Qi2017Loss} of the original GAN to ensure training stability. Wasserstein GAN (WGAN) is one of the most dramatic attempts to handle this problem \cite{arjovsky2017wasserstein}. Arjovsky \emph{et al.} regarded the mini-max game as minimizing the Wasserstein distance between the two distributions and replaced the original loss function of GAN by the Wasserstein distance. Their work significantly improved the stability of GAN training while maintaining the generation ability of GANs. In addition, WGAN requires no extra sophisticated network designation and can be easily applied to the generation of different signals, such as EEG. Based on WGAN, Gulrajani \emph{et al.} proposed using a gradient penalty in the training, which improved the performance of WGAN \cite{gulrajani2017improved}.



\subsection{Data Augmentation}
Data augmentation aims at generating new data of the given dataset by applying transformations to the original data while preserving the label \cite{artData}. This method is commonly applied to reduce overfitting and improve classification performance \cite{Krizhevsky2012ImageNet} since the generated data have a similar data distribution to the original data and can be used to increase the quantity of training data. In the field of image classification with small data size, this technique has been successfully adopted \cite{Perez2017The}. It is common to generate additional images by applying different distortions, scaling, or moving window/pixel shifts to the real images \cite{simard2003best}. A similar technique has also been adopted to generate EEG signals. Krell and Su proposed rotational distortions that were similar to affine/rotational distortions of images to generate artificial EEG signals \cite{Krell2017Rotational}. Lotte generated artificial EEG trials by the relevant combinations and distortions of the original trials \cite{Lotte2015Signal}. Wang \emph{et al.} generated EEG features by directly adding different Gaussian noises to the original feature and applied deep neural networks to verify the effect \cite{Fang2018Data}. All of the abovementioned methods reported that the performance of the classifiers was improved by data augmentation.

Some pioneering works have focused on augmenting data by GANs, which demonstrated great generative ability. Zheng \emph{et al.} adopted DCGAN to generate images and used artificial images for person reidentification tasks \cite{zheng2017unlabeled}. Their results presented the feasibility of GAN-based data augmentation. They also reported that the classifier was less prone to overfitting by adding generated training samples. By applying a CycleGAN to augment the training dataset, Zhu \emph{et al.} improved the classification accuracy of the emotion recognition task based on images \cite{zhu2018emotion}. For EEG signal generation, Hartmann \emph{et al.} proposed EEG-GAN to generate raw EEG signals \cite{Hartmann2018EEG}. In their work, they presented a series of evaluation metrics to demonstrate the potential for GANs to generate EEG data. However, they did not report the performance of the classifier when adding the generated EEG data to the training dataset. In our previous work, we extended the GAN-based augmentation method to EEG-based emotion recognition \cite{luo2018embc}. The experimental results demonstrated the efficiency of our data augmentation method for EEG-based emotion recognition.

\subsection{Deep Neural Networks}

Deep neural networks have shown great success in recent decades. In 2006, Hinton \emph{et al.} proposed an effective method that enabled deep autoencoder networks to learn low-dimensional codes and solved the problem of gradient disappearance \cite{hinton2006reducing}. In the field of computer vision, ImageNet, published in 2009, enables the development and application of large-scale data-driven machine learning methods.
Krizhevsky \emph{et al.} trained a large, deep convolutional neural network called AlexNet using ReLU and a regularization method called `dropout', and acquired inspiring results. Based on the work of Krizhevsky, VGG applied smaller filters and explored the impact of the depth of convolutional neural networks (CNNs) on image recognition accuracy \cite{simonyan2014very}. Except for the field of computer vision, deep neural networks have also been widely applied to the fields of natural language processing \cite{mikolov2013efficient} and speech recognition \cite{dahl2011context}.

Although deep neural networks obtain exciting results in many tasks, they still suffer from problems such as the curse of depth. It is difficult to train a neural network effectively with too many layers. To solve this problem, He \emph{et al.} proposed a residual learning framework called Resnet \cite{he2016deep}, which had shortcuts between layers to transform the information. Inspired by this, we apply the deep neural network (DNN) with shortcut layers as one of our classifiers.

\section{Method}

\begin{figure}[tp]
\centering
\includegraphics[width=0.49\textwidth]{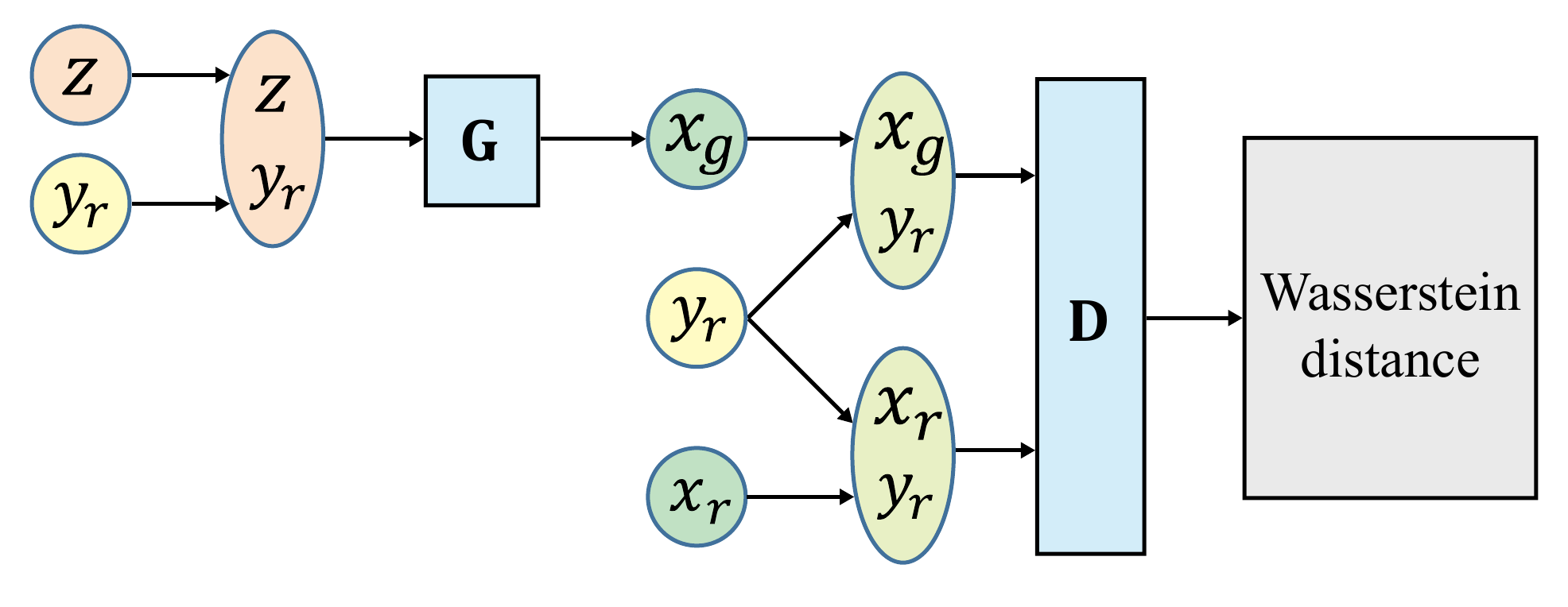}
\caption{The network of cWGAN. Here, $x_r, y_r, x_g, z$, G, and D represent one real sample, real label, generated sample, noise, generator, and discriminator, respectively.}
\end{figure}

In this section, we first give a brief introduction to VAE and WGAN. Then, we present our three deep generative models, cWGAN, sVAE, and sWGAN. Next, we describe three conventional data augmentation methods, cVAE, Gau, and RDA. Finally, we briefly describe DNN with shortcut layers.
\subsection{VAE}
The VAE is a latent variable generative model that consists of an encoder and a decoder. This model combines variational inference with the conventional autoencoder framework. The encoder encodes $x$ into a latent representation space $z$, where $x$ represents a real datapoint and has weights and biases $\lambda$. We denote the encoder $q_\lambda(z|x)$. The decoder outputs the probability distribution of real data given the latent representation $z$. It has weights and biases $\phi$, which is denoted by $p_\phi(x|z)$.

The generative model aims to maximize the probability of each $x$ in the training set according to
\begin{equation}
\begin{split}
p(x) = \int p(x|z)p(z)dx.
\end{split}
\end{equation}
However, this integral requires exponential time to compute. In practice, $p(x|z)$ will be nearly zero for most $z$, which contributes almost nothing to estimate $p(x)$. The VAE attempts to sample $z$, which is likely to produce $x$, by approximating the posterior $p(z|x)$ with $q_\lambda(z|x)$. It uses the Kullback-Leibler (KL) divergence, which measures the distance between two distributions:
\begin{equation}
\begin{split}
KL(q_\lambda(z|x)||p(z|x)) &= \mathbb{E}_{z\sim q}[\log(q_\lambda(z|x)) - \log(p(z|x))]\\&= \mathbb{E}_{z\sim q}[\log(q_\lambda(z|x)) - \log(p(x|z)) \\&- \log(p(z))] + \log(p(x)).
\end{split}
\end{equation}
The goal of KL divergence is to find the parameter $\lambda$ to minimize this divergence. However, it is still impossible to compute the KL divergence directly since $p(x)$ appears in the formula, which is intractable as mentioned above. We can define the following function:
\begin{equation}
\begin{split}
ELBO = &-\mathbb{E}_{z\sim q}[\log(q_\lambda(z|x)) - \log(p(x|z))\\& - \log(p(z))],
\end{split}
\end{equation}
where $ELBO$ represents the evidence lower bound. Combining Eq. (3) with the KL divergence and rewrite, $p(x)$ can be rewritten as
\begin{equation}
\begin{split}
\log(p(x)) = ELBO + KL(q_\lambda(z|x)||p(z|x)).
\end{split}
\end{equation}

Note that the KL divergence is always greater than or equal to zero according to Jensen's inequality. Therefore, minimizing the KL divergence is equivalent to maximizing $ELBO$.

Now, we can decompose the $ELBO$ into a sum where each term depends on a single datapoint since no datapoint shares its latent $z$ with another datapoint in VAE. We can write the $ELBO_i$ for a single datapoint $i$ (the $i^{th}$ datapoint) as
\begin{equation}
\begin{split}
ELBO_i &= -\mathbb{E}_{z\sim q}[\log(q_\lambda(z|x_i)) - \log(p(x_i)|z)) \\&- \log(p(z))] + \log(p(x_i)) \\&= \mathbb{E}_{z\sim q}[\log(p_\lambda(x_i|z))] - KL(q_\lambda(z|x_i)||p(z)),
\end{split}
\end{equation}
where the first term is the expected log-likelihood and the second term is the negative KL divergence between the encoder distributions $q_\lambda(z|x_i)$ and $p(z)$. The first term forces the decoder to learn to reconstruct the data from latent representation, and poor reconstruction results in a large cost in this loss function. The second term can also be viewed as a regularizer, which measures how much information is lost when using $q_\lambda(z|x_i)$ to represent $p(z)$. The encoder receives a penalty if it outputs latent representations $z$ that are different from those from $p(z)$. This term maintains the diversity of the latent representation.

In VAE, the choice of $p(x|z)$ is often a Gaussian distribution. Then, the first term of $ELBO_i$ can also be viewed as the reconstruction loss. The VAE assumes $p(z)=N(0, I)$ and $q_\lambda(z|x_i)=N(\mu(x_i), \Sigma(x_i))$, where $N$ represents a Gaussian distribution. Therefore, the second term of $ELBO_i$ can be formalized as:
\begin{equation}
\begin{split}
KL(q_\lambda(z|x_i)||p(z)) &= KL(N(\mu(x_i), \Sigma(x_i))||N(0, I)) \\&=\frac 12(tr(\Sigma(x_i))+\mu(x_i)^T\mu(x_i)\\&-k-\log(det(\Sigma(x_i)))),
\end{split}
\end{equation}
where $k$ is the dimension of the Gaussian distribution and $tr(x_i)$ is the trace function. We define $\Sigma(x_i)$ as a diagonal matrix, so the formula can be rewritten as
\begin{equation}
\begin{split}
KL(q_\lambda(z|x_i)||p(z)) &= \frac12\sum_k[\Sigma(x_i) + \mu^2(x_i) - 1 - log\Sigma(x_i)].
\end{split}
\end{equation}
In practice, we use $log\Sigma(x_i)$ instead of $\Sigma(x_i)$ since it is more numerically stable to take the exponent. Hence, the final goal of VAE is
\begin{equation}
\begin{split}
\mathop {max\,}\limits_{\lambda, \phi} ELBO &= \sum_i\,ELBO_i \\&= \sum_i\,\sum_k\,[(x_i - \hat{x}_i)^2 + \frac12(\Sigma(x_i) + \mu^2(x_i) \\& - 1 - log\Sigma(x_i))],
\end{split}
\end{equation}
where $\hat{x}_i$ is the reconstructed data, and $\mu(x_i)$ and $log\Sigma(x_i)$ are both calculated by the neural network.

\begin{figure}[tp]
\centering
\includegraphics[width=0.49\textwidth]{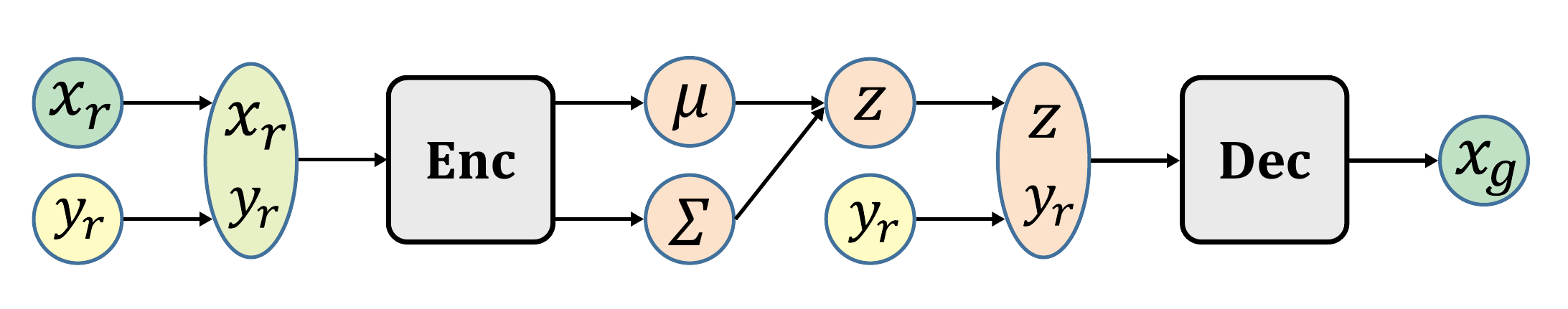}
\caption{The network of cVAE. Here, $x_r, y_r, x_g, \mu, \Sigma, z$, Enc, and Dec represent one real sample, real label, generated sample, mean value, standard deviation, resampled noise, encoder, and decoder, respectively.}
\end{figure}

\subsection{WGAN}
A typical GAN consists of two competing parts, which are both parameterized as deep neural networks. A generator $G$ produces synthetic data given a noise variable input, while a discriminator $D$ attempts to identify whether a sample comes from the real data distribution $X_r$ or the generated data distribution $X_g$. In other words, the discriminator is trained to estimate the probability of a given sample from the real data distribution. The generator is optimized to trick the discriminator to offer a high probability for the generated data. The two parts are optimized simultaneously to find a Nash equilibrium. More formally, the procedure can be expressed as a mini-max function:
\begin{equation}
\begin{split}
   \mathop {\min }\limits_{\theta_G} \mathop {\max }\limits_{\theta_D} L(X_r, X_g) &=
   \mathbb{E}_{x_r\sim X_r}[\log(D(x_r))] \\&+ \mathbb{E}_{z\sim Z}[\log(1-D(G(z)))]\\ &=
   \mathbb{E}_{x_r\sim X_r}[\log(D(x_r))] \\&+  \mathbb{E}_{x_g\sim X_g}[\log(1-D(x_g))],
\end{split}
\end{equation}
where $\theta_g$ and $\theta_d$ represent the parameters of the generator and discriminator, respectively, and $Z$ can be a uniform noise distribution or a Gaussian noise distribution.

The function is optimized in two steps: (i) Maximize it by fixing $G$ and $X_g$, and obtain the optimum of $D$; (ii) Minimize the function by the optional $D$, and then minimize the Jensen-Shannon divergence between $X_r$ and $X_g$. The game achieves equilibrium if and only if $X_r = X_g$.

Although GAN has shown great success in realistic data generation, it suffers from some major problems, such as nonconvergence, mode collapse and diminished gradient. Researchers believed that the Jensen-Shannon divergence could lead to vanishing gradients, which was the main reason for the GAN's instability. In real-world tasks such as image generation, the distribution of real images always lies in low-dimensional manifolds, and the distribution of generated images also rests in low-dimensional manifolds. The two distributions are almost certainly disjoint and have no overlaps. In this situation, the Jensen-Shannon divergence between the two distributions is a fixed number, which cannot provide useful gradients for GAN training.

Arjovsky \emph{et al.} \cite{arjovsky2017wasserstein} adopted the Wasserstein distance, which is also called the earth mover's distance (EM distance), in GAN training to solve the instability problem. The distance formula for the continuous probability domain is
\begin{equation}
   W(X_{r},X_{g}) = \inf_{\gamma \sim \Pi(X_{r},X_{g})}\mathbb{E}_{(x_r,x_g)\sim\gamma}[||x_r-x_g||],
\end{equation}
where $\Pi(X_r, X_g)$ is the set of all possible joint probability distributions between $X_r$ and $X_g$. For the Wasserstein distance, even if the two distributions have no overlaps, it can still provide useful and
smooth gradients for GAN training. However, it is difficult to implement the infimum of Eq. (2).
An alternative method for calculating the Wasserstein distance in reality is to apply its Kantorovich-Rubinstein
duality:
\begin{equation}
\begin{split}
   W(X_{r},X_{g}) = \frac{1}{K} \sup_{||f||_{L} \le K} &\mathbb{E}_{x_r \sim X_{r}}[f(x_r)]\\
   - &\mathbb{E}_{x_g \sim X_{g}}[f(x_g)],
\end{split}
\end{equation}
where $f$ denotes the set of 1-Lipschitz functions and $K$ is a constant number.
In realistic implementations, $f$ is replaced by discriminator $D$ and $||f||_{L} \le K$ is replaced by $||D||_{L} \le 1$.

There are many methods for realizing the 1-Lipschitz constraint in WGAN. One possible method is to
restrict the parameters of the discriminator in a limited range, such as -0.1 to 0.1. However, this weight-clipping
method will introduce some problems. The model may produce poor quality data and does not converge since
clipping reduces the capacity of the model. Another method is to use gradient penalty \cite{gulrajani2017improved}. In this method, an extra penalty term is added to the loss function:
\begin{equation}
\begin{split}
   \mathop {\min }\limits_{\theta_G} \mathop {\max }\limits_{\theta_D} L(X_r, X_g) &=  \mathbb{E}_{x_r \sim X_r}[D(x_r)] \\
   & - \mathbb{E}_{x_g \sim X_g}[D(x_g)] \\
   & - \lambda\mathbb{E}_{\hat{x} \sim \hat{X}}[(||\nabla_{\hat{x}} D(\hat{x})||_{2}-1)^{2}],
\end{split}
\end{equation}
where $\lambda$ is a hyperparameter controlling the trade-off between the original objective and
gradient penalty, and $\hat{x}$ denotes the data points sampled from the straight line
between the real distribution $X_r$ and the generated distribution $X_g$:
\begin{equation}
	\hat{x} = \alpha x_r + (1 - \alpha) x_g, \alpha \sim U[0, 1], x_r \sim X_r, x_g \sim X_g.
\end{equation}

\subsection{cWGAN}
In this paper, we propose the cWGAN and apply it to EEG-based emotion recognition. As shown in Fig. 3, we can generate data with specified categories by using cWGAN. This method is based on the gradient penalty version of WGAN. The cWGAN is formulated as  
\begin{equation}
\begin{split}
   &\mathop {\max }\limits_{\theta_D} L(X_r, X_g, Y_r) \\&= \mathbb{E}_{x_r \sim X_r, y_r \sim Y_r}[D(x_r|y_r)] \\&- \mathbb{E}_{x_g \sim X_g, y_r \sim Y_r}[D(x_g|y_r)] \\
   \\&- \lambda\mathbb{E}_{\hat{x} \sim \hat{X}, y_r \sim Y_r}[(||\nabla_{\hat{x}|y_r} D(\hat{x}|y_r)||_{2}-1)^{2}],
\end{split}
\end{equation}

\begin{equation}
\mathop {\min }\limits_{\theta_G} L(X_g, Y_r) = - \mathbb{E}_{x_g \sim X_g, y_r \sim Y_r}[D(x_g|y_r)],
\end{equation}
where $Y_r$ represents the category distribution of the real data, and $\hat{x}$ is defined in Eq. (13). In this work, $\lambda$ is set to 10 because we find that this value could make the training procedure more stable in our preliminary experiment. The last term in Eq. (14) penalizes the model if the gradient norm moves away from its target norm. The Lipschitz constraint is realized, and the model almost loses no capacity. Thus, cWGAN can generate data with high quality and converge quickly. Since the discriminator loss (\emph{D}-loss) is the
The Wasserstein distance between the two conditional distributions can represent the
training procedure for cWGAN.

\newenvironment{shrinkeq}[1]
{ \bgroup
  \addtolength\abovedisplayshortskip{#1}
  \addtolength\abovedisplayskip{#1}
  \addtolength\belowdisplayshortskip{#1}
  \addtolength\belowdisplayskip{#1}}
{\egroup\ignorespacesafterend}
\begin{shrinkeq}{-1.5ex}
\renewcommand{\algorithmicrequire}{\textbf{Input:}}
\renewcommand{\algorithmicensure}{\textbf{Output:}}
\begin{algorithm*}[htb]
\caption{The work flow of sVAE and sWGAN}
\begin{algorithmic}[1]
\Require
 Real dataset $X_r =\{ x_r^i \}_{i=1}^{m}$ and corresponding labels $Y_r =\{ y_r^i \}_{i=1}^{m}$ and thre\_hold
\Ensure
Generated data $X_g =\{ x_g^i \}_{i=1}^{n}$ and corresponding labels $Y_g =\{ y_g^i \}_{i=1}^{n}$
\State $X_g, Y_g = Null, Null$
\Repeat
\State $X_{all\_g}$ = sVAE($X_r$, noise) or $X_{all\_g}$ = sWGAN($X_r$, noise)
\State $X_{tr}, Y_{tr} = X_r \cup X_g, Y_r \cup Y_g$
\State model = classifier\_train($X_{tr}$, $Y_{tr}$)
\State $Y_{all\_g}$, class\_conf = classifier\_test(model, $X_{all\_g}$)
\For{$i$ in $X_{all\_g}$}
\If{class\_conf[i] $>$ thre\_hold}
$X_g, Y_g = X_g \cup X_{all\_g}[i], \, Y_g \cup Y_{all\_g}[i]$
\EndIf
\EndFor
\Until len($X_g$) == n
\State \Return $X_g, \, Y_g$
\end{algorithmic}
\end{algorithm*}
\label{alg:1}
\end{shrinkeq}

\subsection{sVAE and sWGAN}
In cWGAN, we append all of the generated data to the training dataset. Here, we consider another strategy in which partially generated data are adopted to enlarge the training dataset in sVAE and sWGAN. These two methods are based on the observation that the generated data have different qualities, and only generated data with high quality are selected as new training data. This procedure has two steps: a) we generate some data by VAE or WGAN; b) we choose the generated data with high quality to enlarge the dataset. We repeat the above two steps until we obtain enough training data.

In this work, we use the classification confidence to examine data quality. We first train a classifier with the original training dataset and then use the trained classifier to classify the generated data, and only data with high classification confidence (higher than the $thre\_hold$) are appended to the training dataset. We train a new classifier with the appended dataset and repeat the two steps mentioned above until we have enough generated data. We present the algorithm in Algorithm \ref{alg:1}.

\subsection{cVAE}
In this method, we aim to generate data with the specified category. As shown in Fig. 4, to control the generated category, an extra label is added to the encoder and decoder. We first feed the training data point and the corresponding label to the encoder, then we concatenate the hidden representation with the corresponding label and feed it to the decoder to train the network. Then, we can generate data with the specified label by feeding the decoder with the noise sampled from the Gaussian distribution and the assigned label. Therefore, the cVAE \cite{kingma2014auto-encoding} can be formulated as
\begin{equation}
\begin{split}
&\mathop {max\,}\limits_{\lambda, \phi} ELBO \\&= \sum_i\,ELBO_i \\&= \sum_i\,\sum_k\,[(x_i - \hat{x}_i)^2 \\&+ \frac12(\Sigma(x_i|y_i) + \mu^2(x_i|y_i) - 1 - log\Sigma(x_i|y_i))].
\end{split}
\end{equation}

\subsection{Gaussian noise}
One of the straightforward augmentation methods is adding Gaussian noise (Gau) to the original training data, whose probability density function obeys a Gaussian distribution:
\begin{equation}
 p_G(z)=\frac{1}{\sigma \sqrt{2\pi}}e^{-\frac{(z-\mu )^2}{2\sigma^2}},
\end{equation}
where $z$ is a random variable, $\mu$ means expectation and $\sigma$ is the standard deviation. In our experiment, $\mu$ is set to 0 and $\sigma$ is set to 0.001. Intuitively, more training data can be generated while preserving the characteristics of the original data by adding Gaussian noise.

\subsection{Rotational Data Augmentation}
Rotational data augmentation (RDA) was proposed by Krell and Su \cite{Krell2017Rotational}, which
aims to create data with strong spatial robustness, since there might be
spatial shifts of EEG caps within sessions and between sessions during the
experiments. To address this problem, RDA generates artificial
data associated with the electrodes' new positions by adding rotations on three coordinates.
According to their result \cite{Krell2017Rotational}, augmentation around the \emph{y}-axis and \emph{z}-axis increased the performance, especially around the \emph{z}-axis. Therefore, we choose to perform the rotations
around the \emph{z}-axis. Specifically, we set an angle between $12^\circ$ and $24^\circ$ over all subjects and calculate the new data by interpolation based on radial basis functions.

\subsection{Classifier}
In this paper, we implement two kinds of classifiers: SVMs and deep neural networks.

The deep neural network is a neural network with multiple hidden layers. Here, we randomly add some residual functions between two layers. The idea of the residual function is borrowed from Resnet \cite{he2016deep}. The residual function is a way to avoid the problem of vanishing gradients, and it does this by using shortcuts to jump over some layers. Because the numbers of different nodes are different, the dimensions of input and output are different. Therefore, we can use a linear projection to match the dimensions. This function can be expressed as follows:
\begin{equation}
	y = F(x)+W_sx,
\end{equation}
where $x$ and $y$ are the input and output vectors, respectively, $F(x)=W_2\sigma(W_1x)$, which means two fully connected layers and a ReLU function, $\sigma$, between them, and $W_s$ is the linear projection to change the dimension. The output should go through another ReLU function before it is passed to the next layer.

\section{Experimental Settings}
In this section, we describe the details about the two datasets, data preprocessing, performance evaluation, and hyperparameters.
\subsection{Dataset Description}
The SEED dataset \cite{zheng2015investigating} contains the EEG signals of 15 participants.
They were required to watch 15 well-prepared video clips that can elicit exactly one of the three kinds of emotion: positive, neutral, and negative. The criteria of film clip selection ensure that
Each clip is well-edited to create coherent emotion eliciting and maximizing emotional meanings.
In addition, each clip can explicitly elicit one exact kind of emotion, and the time of the clips is enough but not too long to elicit the participants' corresponding emotion sufficiently.
The order of presentation is arranged so that two film clips targeting the same emotion are not shown consecutively. Each participant took part in the experiment three times with an interval of at least 7 days. The signals were sampled at a rate of 1,000 Hz with an ESI NeuroScan System from a 62-electrode headset.

The DEAP dataset\cite{koelstra2012deap} contains the EEG and peripheral physiological signals of 32 participants as they watched 40 one-minute-long excerpts of music videos. The music videos were selected from 120 one-minute extracts of music videos rated from an online self-assessment by
14-16 volunteers based on valence, dominance, arousal, like, and familiarity. Valence, arousal, dominance and like were rated directly after each trial on a continuous 9-point scale using a standard mouse self-assessment. The signals were sampled at 512 Hz with 48 channels.
(32 EEG channels, 12 peripheral physiological channels including
galvanic skin response and temperature, 3 unused channels and 1 status channel).
The signals from EEG channels are sampled according to an international
10-20 system.

\subsection{Data Preprocessing}
Previous works have shown that the DE feature of EEG signals is efficient for EEG-based emotion recognition \cite{zheng2015investigating,zheng2017identifying, yang2018eeg-based}. Therefore, we generate DE features to augment the datasets. We also generate the PSD feature, which is a conventional feature for EEG-based emotion recognition, to verify our method. Since both the SEED and DEAP datasets have been preprocessed, we use the short-time Fourier transform (STFT) with a 1-s-long nonoverlapping overleaping Hanning window to extract the PSD feature of the EEG signal from the two datasets directly. For the Gaussian distribution, the DE feature is defined as
\begin{equation}
\begin{split}
    h(X) = &-\int_{-\infty}^{\infty} \frac{1}{\sqrt{2\pi\sigma^2}}\exp\frac{(x-\mu)^2}{2\sigma^2}\log\frac{1}{\sqrt{2\pi\sigma^2}} \\
    &exp\frac{(x-\mu)^2}{2\sigma^2}dx = \frac{1}{2}\log2\pi e\sigma^2,
\end{split}
\end{equation}
where $X$ represents the Gaussian distribution $N(\mu, \sigma^2)$, and $\pi$ and $e$ are constants. Shi \emph{et al.} \cite{Shi2013Differential} demonstrated that the value of DE is equal to the logarithmic spectral energy for a fixed-length EEG sequence in a certain band. According to their result, we extracted the DE feature from the preprocessed EEG signal of the two datasets.

Considering the dynamic characteristics of EEG-based emotion recognition tasks, we employ the linear dynamic system approach to filter the PSD and DE features, which has also been adopted in previous works \cite{zheng2015investigating,zheng2017identifying}.

PSD and DE features are extracted from five frequency bands: $\delta$: 1-3 Hz, $\theta$: 4-7 Hz, $\alpha$: 8-13 Hz, $\beta$: 14-30 Hz, and $\gamma$: 31-50 Hz for the SEED dataset \cite{zheng2015investigating}. Therefore, both of these features have 310 dimensions (62 channels $\times$ 5 frequency bands). For each experiment, there were 3,394 labeled samples. In this work, we viewed the SEED dataset as a three-category classification problem.

We also extracted PSD and DE features for the DEAP dataset. Since the $\delta$ band was filtered in this dataset, we only computed the two features of four frequency bands: $\theta$, $\alpha$, $\beta$, and $\gamma$. In this time, both features had 128 dimensions (32 channels $\times$ 4 frequency bands). Each experiment had 2,400 labeled samples. Here, we adopted a four-category emotion model using valence and arousal values: high valence (level $>$ 5) and high arousal (level $>$ 5), high valence (level $>$ 5) and low arousal (level $\le$ 5), low valence (level $\le$ 5) and high arousal (level $>$ 5), and low valence (level $>$ 5) and high arousal (level $>$ 5).

\subsection{Evaluation Details}
We adopted 5-fold cross-validation for each experiment on the two datasets. For each experiment, we trained 5 recognition models and computed the average recognition accuracy of the five models as the recognition accuracy of the experiment. Each model had the same hyperparameters. We regarded the average accuracy of all experiments as the final accuracy. For the SEED dataset, there were 45 experiments and nearly 678 samples for each fold. For the DEAP dataset, there were 32 experiments and 480 samples for each fold.

\subsection{Hyperparameter Details}
In the SVM classifier, we used the linear kernel. The parameter $c$ was searched from $2^{-10}$ to $2^{10}$ to find the optimal value.

We performed a random search on the learning rate, number of network layers and size of batches of classifier of deep neural network. The learning rate was randomly selected from 0.0005, 0.0001, 0.00005 and 0.00001 with the Adam optimizer. The number of layers was searched from 4 to 8. The size of batches was randomly selected from 128, 256 and 512. Both networks with residual functions and without residual functions were searched. For the network with residual functions, the residual functions were applied every two layers. The input dimension was determined by the corresponding input feature, and the dimension of the output label was 3 for the SEED dataset and 4 for the DEAP dataset. The number of hidden nodes for each layer was randomly searched. The ReLU activation function was used for all hidden layers. We normalize PSD and DE features before feeding them to the networks.

\begin{table*}[ht]
\footnotesize
\caption{Mean accuracies/standard deviations of SVMs on the SEED dataset and appending datasets using PSD feature generated by different methods. `$\uparrow$' represents the maximize accuracy improvement of different methods and has the same meaning in tables \ref{tab:2}, \ref{tab:3}, and \ref{tab:4}.}
\begin{center}
\renewcommand\arraystretch{1.1}
\setlength{\tabcolsep}{0.9mm}{
\begin{tabular}{lllllllllll}
\toprule[1.2pt]
\diagbox[width=9.5em,height=2.7em]{methods}{No. of append} & 0 &  200 & 500 & 1000 & 3000 & 5000 & 10000 & 15000 & 20000 & $\uparrow$\\
\midrule
cVAE + SVM
&60.3/15.9&62.7/15.7&62.8/15.4&63.4/14.6&\textbf{63.4}/14.8&63.3/\textbf{14.3}&62.5/14.8&61.8/14.8&61.6/14.5&3.1\\
Gau + SVM
&60.3/15.9&61.4/15.6&61.7/15.5&61.7/15.7&62.5/15.4&62.5/15.6&\textbf{63.1}/\textbf{15.0}&62.7/15.5&62.8/15.4&2.8\\
RDA + SVM
&60.3/15.9&62.6/15.7&\textbf{63.2}/15.4&62.9/15.8&62.0/15.6&62.1/\textbf{15.2}&61.5/15.8&61.9/15.9&61.1/16.3&2.9\\
\midrule
cWGAN + SVM &60.3/15.9&62.7/15.5&63.6/15.6&63.5/15.6&64.0/15.6&64.4/15.5&65.0/15.6&\textbf{65.2}/\textbf{15.5}&64.9/15.5&4.9\\
sVAE + SVM
&60.3/\textbf{15.9}&62.7/16.9&62.6/16.6&63.3/16.6&62.8/16.9&63.1/16.5&63.4/17.4&\textbf{63.5}/17.2&63.2/17.4&3.2\\
sWGAN + SVM
&60.3/15.9&65.2/\textbf{14.5}&66.0/14.8&66.8/14.9&67.0/14.7&67.0/14.7&67.4/14.8&67.3/15.2&\textbf{67.7}/15.1&\textbf{7.4}\\
\bottomrule[1.2pt]
\end{tabular}}
\end{center}
\label{tab:1}
\end{table*}

\begin{table*}[ht]
\footnotesize
\caption{Mean accuracies/standard deviations of SVMs and deep neural network (DNNs) with shortcut layers on the SEED dataset and appending datasets using DE feature generated by different methods.}
\begin{center}
\renewcommand\arraystretch{1.1}
\setlength{\tabcolsep}{0.9mm}{
\begin{tabular}{lllllllllll}
\toprule[1.2pt]
\diagbox[width=9.5em,height=2.7em]{methods}{No. of append} & 0 & 200 & 500 & 1000 & 3000 & 5000 & 10000 & 15000 & 20000 & $\uparrow$\\
\midrule
cVAE + SVM&84.3/8.7&84.8/8.7&85.2/8.6&\textbf{85.2}/8.6&84.9/8.5&84.9/\textbf{8.5}&84.5/8.9&84.0/8.9&84.0/8.9&0.9\\
cVAE + DNN&83.3/8.2&83.9/9.3&84.9/8.5&86.1/8.0&\textbf{86.5}/\textbf{7.5}&86.1/8.2&85.1/7.8&85.1/8.5&86.1/8.2&3.2\\
Gau + SVM&84.3/8.7&84.6/8.7&84.8/8.6&84.9/8.6&\textbf{85.1}/\textbf{8.5}&85.0/8.6&85.0/8.7&84.8/8.6&84.8/8.5&0.8\\
Gau + DNN&83.3/8.2&85.9/7.6&85.7/7.9&85.0/8.5&85.6/\textbf{7.3}&85.3/8.3&\textbf{86.2}/8.2&84.9/8.5&85.8/7.8&2.9\\	RDA + SVM&84.3/\textbf{8.7}&85.4/9.0&85.5/9.1&85.5/9.0&85.4/8.9&\textbf{85.6}/8.8&84.7/9.1&84.3/9.3&84.3/9.3&1.3\\
RDA + DNN&83.3/\textbf{8.2}&\textbf{85.7}/9.9&83.4/9.5&82.1/9.7&78.2/9.6&77.6/10.9&77.6/10.1&74.6/8.9&75.7/9.1&2.4\\
\midrule
cWGAN + SVM&84.3/8.7&87.0/8.6&87.2/8.5&86.8/8.4&87.0/8.4&87.0/8.5&87.4/8.0&\textbf{87.4}/7.9&87.1/\textbf{7.9}&3.1\\
cWGAN + DNN&83.3/8.2&86.6/7.7&89.2/7.9&89.7/8.3&\textbf{91.6}/\textbf{6.7}&90.9/7.9&90.6/7.9&90.6/8.8&90.7/7.8&8.3\\
sVAE + SVM&84.3/8.7&87.4/7.9&87.5/7.6&\textbf{87.8}/\textbf{7.6}&86.8/8.1&86.1/8.6&85.2/8.7&84.7/8.1&84.5/8.1&3.5\\
sVAE + DNN&83.3/8.2&85.8/8.8&86.8/7.3&\textbf{87.5}/8.6&87.2/6.8&84.1/6.7&84.0/6.5&82.2/\textbf{6.2}&80.4/6.6&4.2\\
sWGAN + SVM&84.3/8.7&87.9/8.4&88.9/8.3&89.7/7.9&90.1/7.6&90.7/7.8&\textbf{90.8}/7.7&90.8/\textbf{7.3}&90.8/7.4&6.5\\
sWGAN + DNN&83.3/8.2&91.4/7.2&91.5/6.4&\textbf{93.5}/5.7&93.5/5.8&93.0/5.8&93.1/\textbf{5.6}&91.7/6.0&92.2/5.7&\textbf{10.2}\\
\bottomrule[1.2pt]
\end{tabular}}
\end{center}
\label{tab:2}
\end{table*}

\begin{table*}[ht]
\footnotesize
\caption{Mean accuracies/standard deviations of SVMs on the DEAP dataset and appending datasets using PSD feature generated by different methods.}
\begin{center}
\renewcommand\arraystretch{1.1}
\setlength{\tabcolsep}{0.9mm}{
\begin{tabular}{lllllllllll}
\toprule[1.2pt]
\diagbox[width=9.5em,height=2.7em]{methods}{No. of append} & 0 & 200 & 500 & 1000 & 3000 & 5000 & 10000 & 15000 & 20000 & $\uparrow$\\
\midrule
cVAE + SVM&42.7/9.6&43.7/9.5&44.5/\textbf{8.7}&44.2/9.3&\textbf{44.9}/8.8&44.6/8.8&44.1/8.9&44.1/9.1&43.4/9.1&2.2\\
Gau + SVM&42.7/9.6&43.2/9.5&43.6/\textbf{9.2}&43.7/9.8&43.9/9.7&\textbf{44.5}/9.3&44.0/9.6&43.9/9.6&43.9/9.5&1.8\\
RDA + SVM&42.7/9.6&42.8/10.0&43.0/9.7&44.1/9.2&44.3/9.2&44.9/\textbf{8.5}&44.7/8.9&44.9/9.0&\textbf{45.2}/8.9&2.5\\
\midrule
cWGAN + SVM&42.7/9.6&44.1/9.6&44.2/9.7&44.8/9.1&44.9/8.8&\textbf{45.0}/8.9&44.9/\textbf{8.7}&44.9/9.0&44.8/9.2&2.3\\
sVAE + SVM&42.7/9.6&44.7/8.4&45.1/\textbf{8.1}&45.6/8.5&45.6/8.3&45.8/8.3&46.1/8.2&\textbf{46.1}/8.5&45.9/8.7&3.4\\
sWGAN + SVM&42.7/\textbf{9.6}&45.8/10.6&45.8/11.0&46.4/10.4&46.7/10.3&46.9/10.4&47.1/10.1&47.4/10.0&\textbf{47.6}/9.9&\textbf{4.9}\\
\bottomrule[1.2pt]
\end{tabular}}
\end{center}
\label{tab:3}
\end{table*}

\begin{table*}[ht]
\footnotesize
\caption{Mean accuracies/standard deviations of SVMs and deep neural networks with shortcut layers on the DEAP dataset and appending datasets using DE feature generated by different methods.}
\begin{center}
\renewcommand\arraystretch{1.1}
\setlength{\tabcolsep}{0.9mm}{
\begin{tabular}{lllllllllll}
\toprule[1.2pt]
\diagbox[width=9.5em,height=2.7em]{methods}{No. of append} & 0 & 200 & 500 & 1000 & 3000 & 5000 & 10000 & 15000 & 20000 & $\uparrow$\\
\midrule
cVAE + SVM&45.4/8.2&46.3/8.1&46.8/8.0&47.2/7.8&47.9/7.7&47.8/7.7&\textbf{48.1}/\textbf{7.6}&48.0/7.7&47.8/7.6&2.7\\
cVAE + DNN&44.9/4.0&\textbf{46.6}/4.4&45.8/3.8&45.5/4.9&45.7/4.6&46.5/4.9&45.7/3.9&45.9/\textbf{3.6}&45.9/4.5&1.7\\
Gau + SVM&45.4/8.2&46.1/\textbf{8.0}&46.0/8.2&\textbf{46.1}/8.1&46.1/8.1&45.9/8.2&46.0/8.2&45.8/8.1&45.8/8.2&0.7\\
Gau + DNN&44.9/4.0&45.9/\textbf{3.7}&\textbf{46.9}/4.2&45.5/4.1&45.5/4.5&46.2/4.6&45.8/4.5&45.6/4.6&46.2/4.5&2.0\\
RDA + SVM&45.4/8.2&45.9/8.2&45.9/8.2&46.1/8.5&\textbf{46.3}/8.1&46.3/8.1&46.1/\textbf{7.8}&46.0/7.8&45.9/7.9&0.9\\
RDA+DNN&44.9/4.0&46.3/4.5&\textbf{46.8}/4.3&46.1/3.7&46.1/4.7&46.0/4.1&45.8/\textbf{3.4}&45.1/3.7&45.9/4.1&1.9\\
\midrule
cWGAN + SVM&45.4/\textbf{8.2}&47.3/8.2&47.9/8.2&48.0/8.3&48.8/8.3&\textbf{48.9}/8.7&48.5/8.4&48.2/8.8&48.0/8.9&3.5\\
cWGAN + DNN&44.9/4.0&45.4/4.4&45.9/\textbf{4.0}&47.2/5.1&47.0/4.4&46.9/4.8&47.1/4.6&\textbf{47.5}/4.5&46.9/4.8&2.6\\
sVAE + SVM&45.4/8.2&47.6/7.1&48.3/7.1&48.2/7.3&48.3/6.8&\textbf{48.4}/\textbf{6.8}&48.1/7.1&48.2/7.0&48.1/7.1&3.0\\
sVAE + DNN&44.9/4.0&47.3/\textbf{3.9}&47.7/4.6&47.6/4.2&47.3/4.2&\textbf{49.3}/5.0&47.7/5.3&47.7/4.4&47.6/6.0&4.4\\
sWGAN + SVM&45.4/8.2&47.6/7.7&47.9/7.5&48.7/7.4&49.6/7.2&49.9/6.8&50.3/7.0&\textbf{50.8}/6.9&50.4/\textbf{6.7}&\textbf{5.4}\\
sWGAN + DNN&44.9/\textbf{4.0}&47.2/4.1&47.7/4.7&47.6/4.4&48.2/4.9&\textbf{49.1}/5.6&47.5/4.6&48.5/5.2&47.6/5.3&4.2\\
\bottomrule[1.2pt]
\end{tabular}}
\end{center}
\label{tab:4}
\end{table*}

\begin{figure*}[ht]
\centering
    \subfloat[]{
            \begin{minipage}[t]{0.45\linewidth}
                \centering
                \includegraphics[height=0.23\textheight,width=1\textwidth]{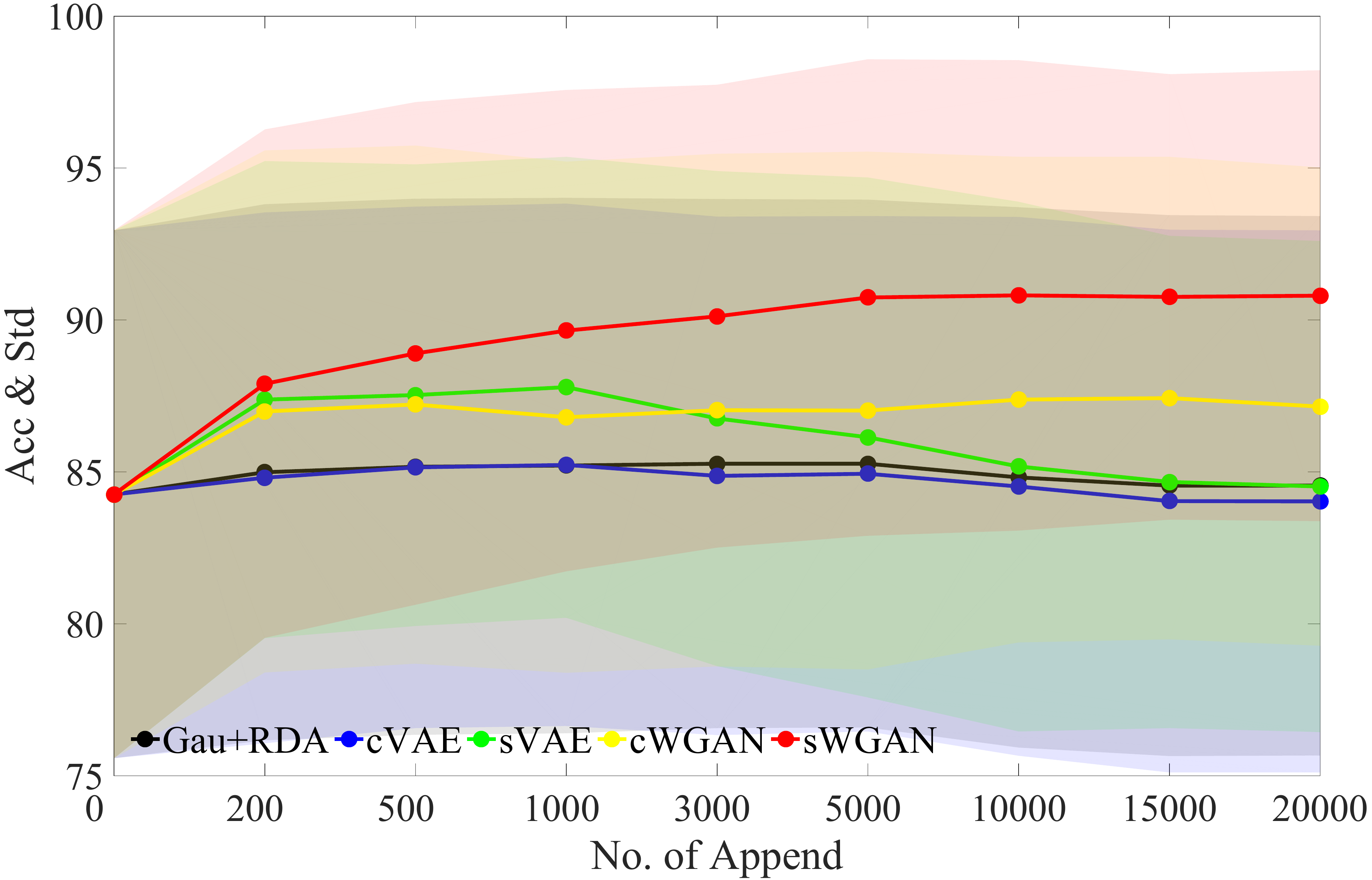}
                \label{fig:4.2}
            \end{minipage}
        }
    \subfloat[]{
            \begin{minipage}[t]{0.45\linewidth}
                \centering
                \includegraphics[height=0.23\textheight,width=1\textwidth]{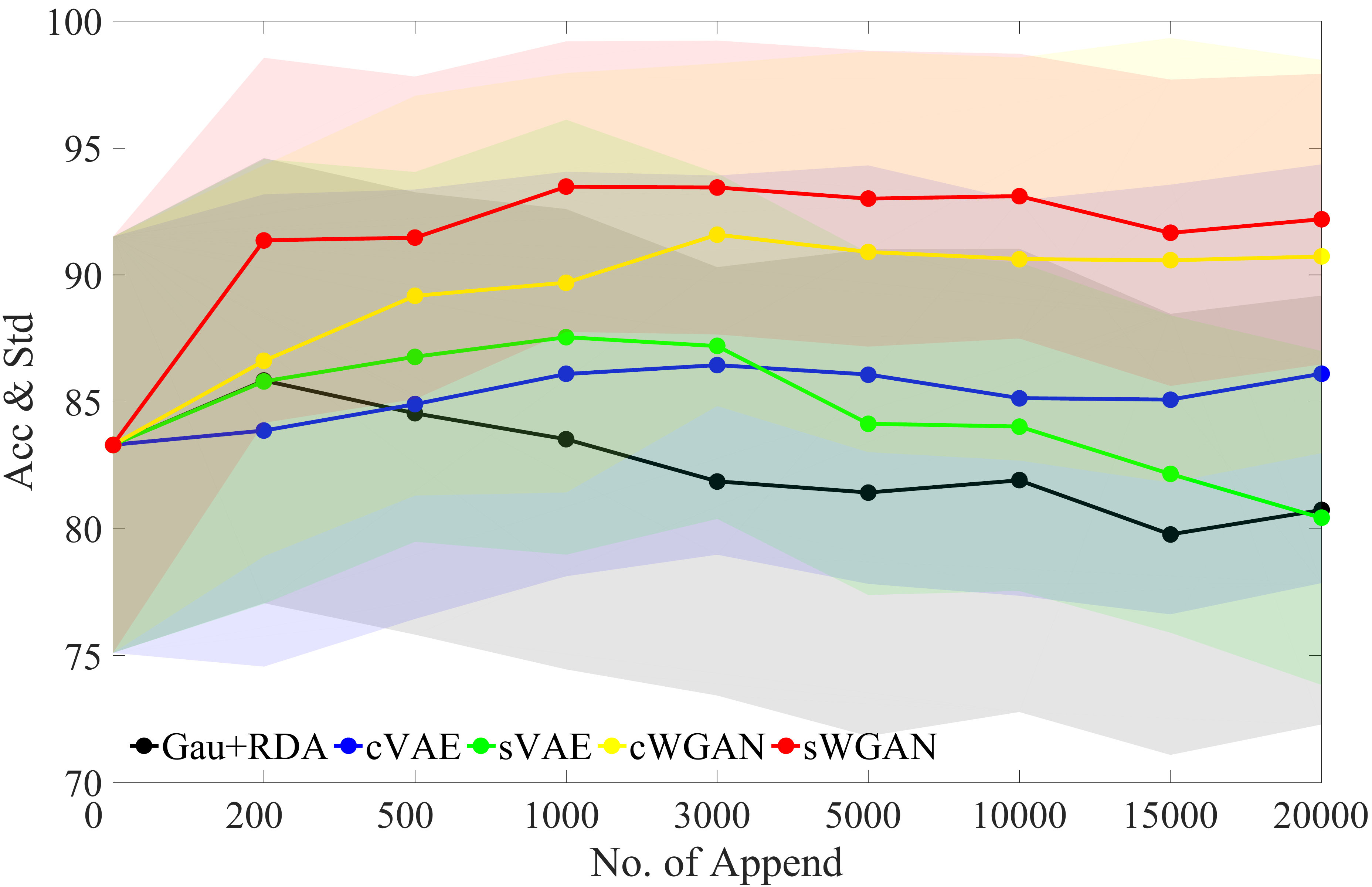}
                \label{fig:4.2}
            \end{minipage}
        }

    \subfloat[]{
            \begin{minipage}[t]{0.45\linewidth}
                \centering
                \includegraphics[height=0.23\textheight,width=1\textwidth]{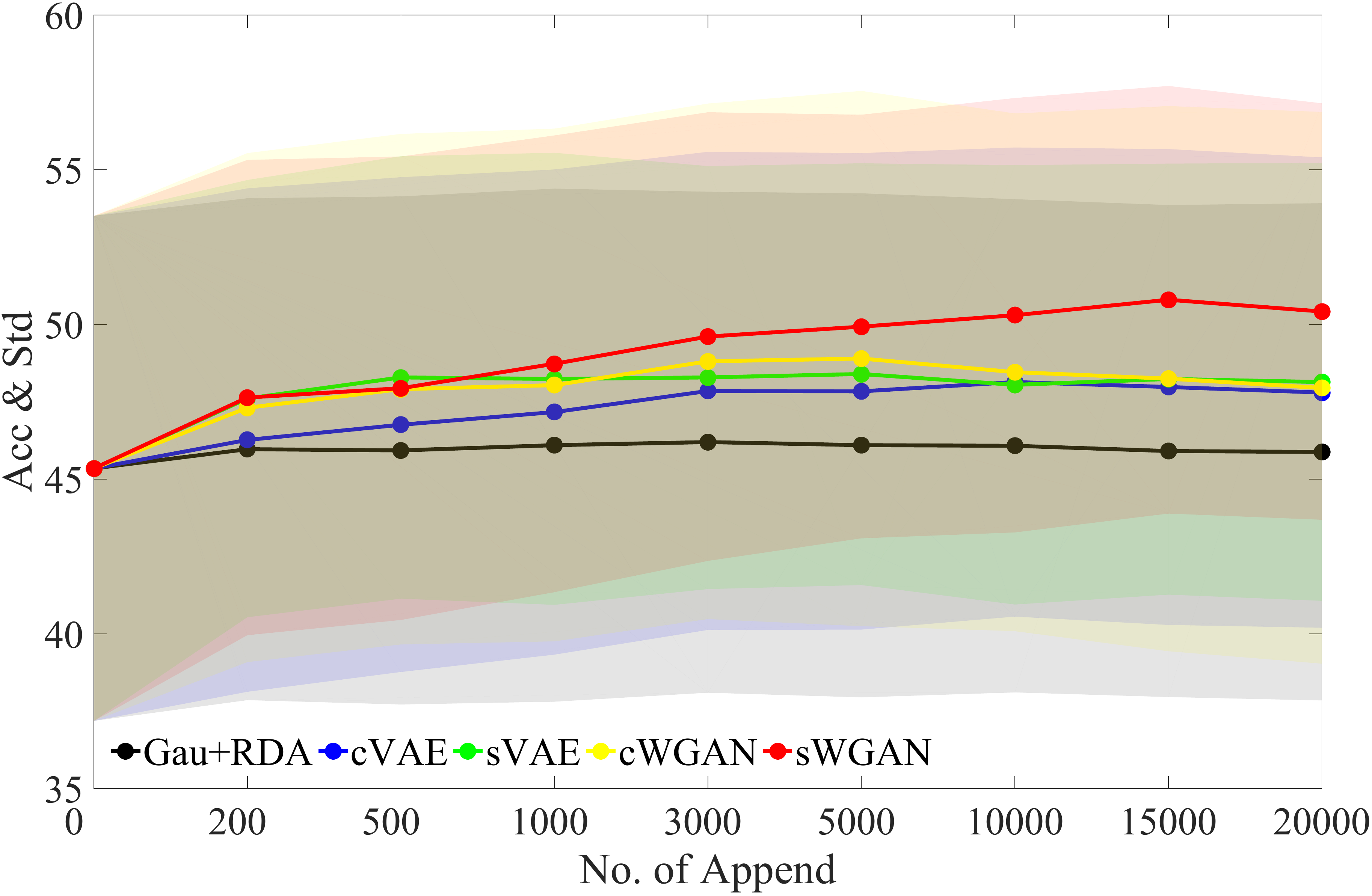}
                \label{fig:4.4}
            \end{minipage}
    }
    \subfloat[]{
            \begin{minipage}[t]{0.45\linewidth}
                \centering
                \includegraphics[height=0.23\textheight,width=1\textwidth]{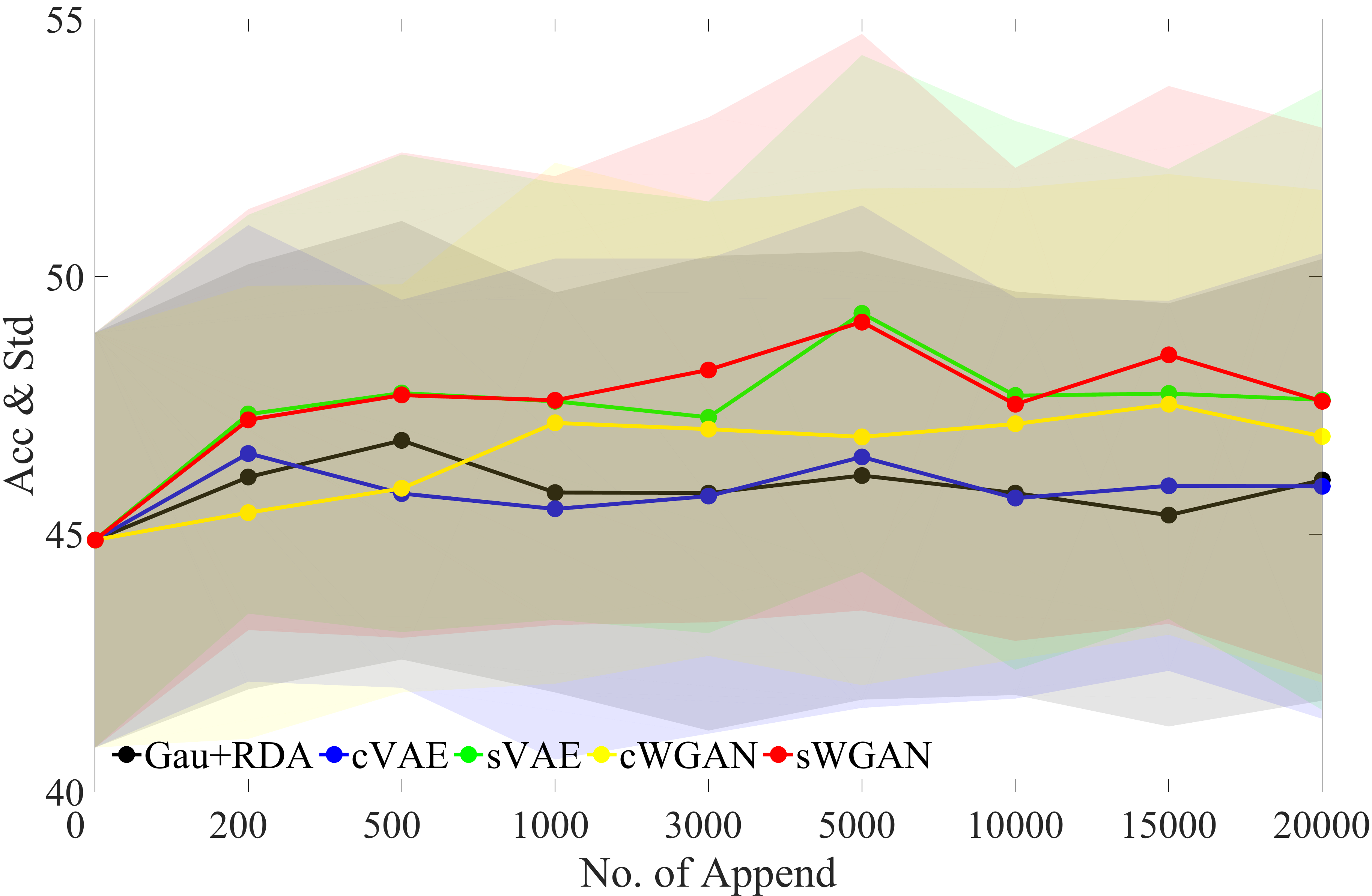}
                \label{fig:4.4}
            \end{minipage}
    }
    \caption{Mean accuracies (Acc) and standard deviations (Std) of different methods on different classifiers and datasets: (a) Acc and Std of SVM on the SEED dataset and appending datasets using DE feature generated by different methods; (b) Acc and Std of DNN on the SEED dataset and appending datasets using DE feature generated by different methods; (c) Acc and Std of SVM on the DEAP dataset and appending datasets using DE feature generated by different methods;(d) Acc and Std of DNN on the DEAP dataset and appending datasets using DE feature generated by different methods.}
    \label{fig:4}
\end{figure*}

\begin{figure*}[ht]
\centering
\includegraphics[width=0.98\textwidth]{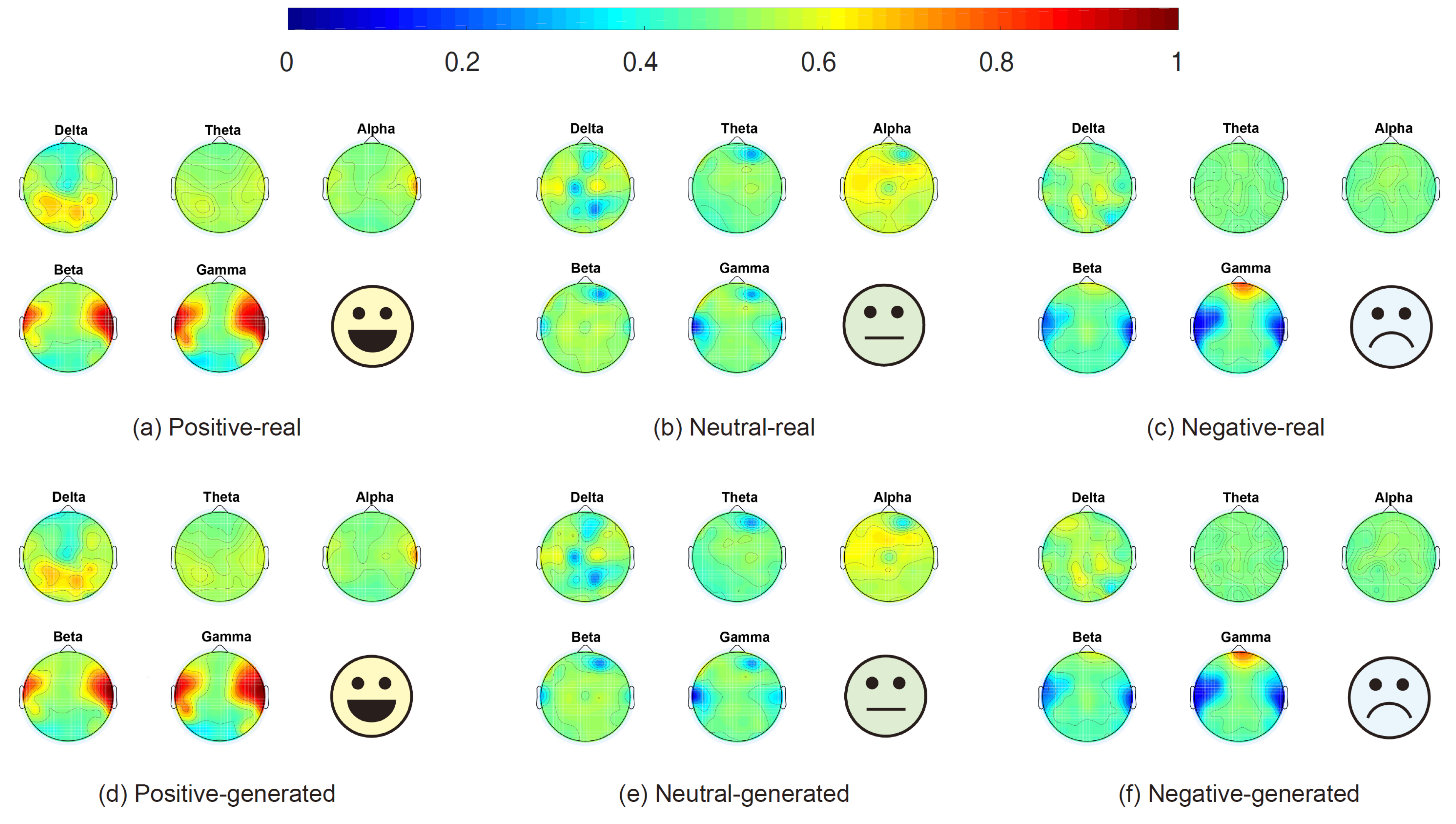}
\caption{Topographic map of the scalp for real and generated DE features (cWGAN) in the SEED dataset.}
\end{figure*}

\section{Experiments and Results}
In this section, we first perform a systematic experimental study to evaluate the effectiveness and generalization ability of our methods. We augmented different EEG-based emotion datasets by different features generated by our methods. We apply different classifiers to evaluate the performances of these generative methods. We also compare our proposed methods with conventional generative methods. Then, we visualized the generated data to show why our proposed methods work. Finally, we discuss the proposed methods.

\subsection{Different Number of Appended Training Data}

We first conducted data augmentation experiments on the SEED dataset and use the SVM as the classifier. Each experiment had 3,394 samples. We generated 0, 200, 500, 1000, 3,000, 5,000, 10,000, 15,000 and 20,000 artificial samples of the two features and added them to the original training datasets. Here, `0' indicates that we only use the original training dataset without data augmentation. We did not generate more samples because we found that most of the experiments reach their peaks before 20,000 samples were appended. The remaining experiments reached their peaks when 20,000 samples were appended. And the $p-values$ between sWGAN method and the conventional methods are all less than 0.01.

We compared the performances of different data augmentation methods when applying the PSD feature, as shown in table \ref{tab:1}. The average accuracy was 60.3\% when we only used the original training set. For conventional methods, cVAE reached its best mean accuracy of 63.4\% when 3,000 samples were appended. Gau reached its optimal performance of 63.1\% when adding 10,000 samples into the original training set. RDA had the best performance of 63.2\% when 500 samples were appended. For our methods, cWGAN achieved its best mean accuracy of 65.2\% when 15,000 samples were appended. When 15,000 samples were appended, sVAE reached its best mean accuracy of 63.5\%. sWGAN achieved its best mean accuracy of 67.7\% when 20,000 samples were appended. According to table \ref{tab:1}, our methods achieved better performance than conventional methods. sWGAN had the best performance among all the methods.

Table \ref{tab:2} illustrates the results of the data augmentation methods for the DE feature. For SVM, the baseline was 84.3\%. For conventional data augmentation methods, cVAE reached its best result of 85.2\% when 1,000 sampled data points were appended. The best accuracy for Gau was 85.1\% with 3,000 augmented data points. The best mean accuracy of RDA was 85.6\% when 5,000 samples were appended. For our methods, cWGAN reached its best mean accuracy of 87.4\% when appending 15,000 samples. For the two selective augmentation methods, sVAE achieved the best mean accuracy of 87.8\% when 1,000 samples were appended, and sWGAN achieved the best mean accuracy of 90.8\% when 10,000 samples were appended.

\subsection{Classification with Deep Neural Networks}

To increase the reliability of the performance comparison of different data augmentation approaches, we also implemented deep neural networks with shortcut layers to build the affective models. Considering that the DE feature is better for the PSD feature in emotion recognition tasks and that the PSD feature had similar improvements in terms of the mean accuracy with the DE feature, we only augmented the training data with the DE features when using the DNN as the classifier. The baseline was 83.3\%. For conventional methods, cVAE, Gau, and RDA reached the best mean accuracy of 86.5\% (3000 samples), 86.2\% (10000 samples), and 85.7\% (200 samples), respectively. For our proposed methods, the best mean accuracy of cWGAN is 91.6\% when we added 3,000 samples. The two selective methods obtained the best mean accuracy of 87.5\% and 93.5\% when we added 1,000 samples, respectively. The results in table \ref{tab:2} demonstrate that our methods had better performance than conventional methods. The sWGAN achieved the best performance for both classifiers.

\subsection{Generated Data with Two Different Features}

For the DEAP dataset, we also used different data augmentation methods to augment PSD and DE features. Each experiment had 2,400 samples. We generated the same number of samples as mentioned above.

Table \ref{tab:3} shows the mean accuracies and standard deviations of PSD data augmentation. The mean accuracy of the 4-category emotion recognition model was 42.7\% when we only used the original training data. For conventional methods, cVAE reached the best mean accuracy of 44.9\% when 3,000 samples were appended. The best performance for Gau was 44.5\% when 5,000 samples were appended. RDA reached the best mean accuracy of 45.2\% when 20,000 samples were appended. For our methods, cWGAN obtained the best mean accuracy of 45.0\% when we added 5,000 generated samples to the original training dataset. sVAE had the best mean accuracy of 46.1\% when 15,000 samples were generated. sWGAN achieved its best mean accuracy of 47.6\% when 20,000 samples were appended. Our methods also showed better performance, and sWGAN had the best performance in terms of accuracy.

Table \ref{tab:4} presents the results of DE data augmentation. For SVM, the baseline was 45.4\%. For conventional methods, cVAE had the best mean accuracy of 48.1\% when 10,000 samples were appended. The best accuracy for Gau was 46.1\% when we appended 1,000 samples. RDA obtained the best mean accuracy of 46.3\% when the number of appended samples was 3,000. For our methods, cWGAN obtained the best mean accuracy of 48.9\% when 5,000 samples were appended. sVAE reached the best mean accuracy of 48.4\% when 5,000 samples were appended, and sWGAN obtained the best mean accuracy of 50.8\% when 15,000 samples were appended.

For DNN, the classification accuracy was 44.9\% when no data augmentation method was applied. For conventional methods, cVAE, Gau, and RDA reached the best mean accuracies of 46.6\% (200 samples), 46.9\% (500 samples), and 46.8\% (500 samples), respectively. For our methods, cWAGN, sVAE, and sWGAN achieved the best mean accuracy of 47.5\% (15,000 samples), 49.3\% (5,000 samples), and 49.1\% (5,000 samples). We also observed that our methods showed better performance than conventional methods. sWGAN had the best mean accuracy when applying SVM as the classifier, while sVAE had the best performance when applying DNN as the classifier.

As we can see from the above results, the DE feature had better mean accuracies than the PSD feature on both datasets, and the standard deviations were smaller. These results were consistent with previous studies \cite{zheng2015investigating,zheng2017identifying}. In addition, compared with conventional methods, our methods were more efficient for improving the performance of emotion recognition models. For the SEED dataset, the mean accuracy improved 10.2\% with DE features when we used sWGAN as the data augmentation method and adopted DNN as the classifier. The DEAP dataset had the highest improvement of 5.4\% in terms of mean accuracy when sWGAN was adopted as the data augmentation method and SVM was used as the classifier. Moreover, the data augmentation methods were more efficient for DNN in most cases.

In addition, we observe that all the data augmentation methods (both ours and conventional methods) reached their peaks, and then their performance decayed when we gradually increased the number of appended samples. However, for our methods, most of the experiments still showed better performance compared with their baselines when appending fewer than 20,000 generated samples. Although experiments with different datasets, features, classifiers, and methods reached different peaks, our results show that the peaks appeared before the training datasets were enlarged 10 times.

As shown in Fig. 5, we plotted the mean accuracies and standard deviations of different methods on different classifiers and datasets. We only shown the results of DE feature because DE feature had better performance than PSD feature and the two features had the similar tendency in terms of mean accuracy. And the results of Gau and RDA were averaged. Compared with the conventional methods, our methods had better mean accuracies in the most experiments. Besides, GAN-based methods (cWGAN and sWGAN) shown better performance than VAE-based (cVAE and sVAE) methods in most cases. Moreover, the selective methods (sVAE and sWGAN) were better than conditional methods (cVAE and cWGAN) in most of the experiments. Specially, sWGAN always had better mean accuracies than cWGAN.

\begin{figure}[ht]
\centering\includegraphics[height=0.28\textheight, width=0.45\textwidth]{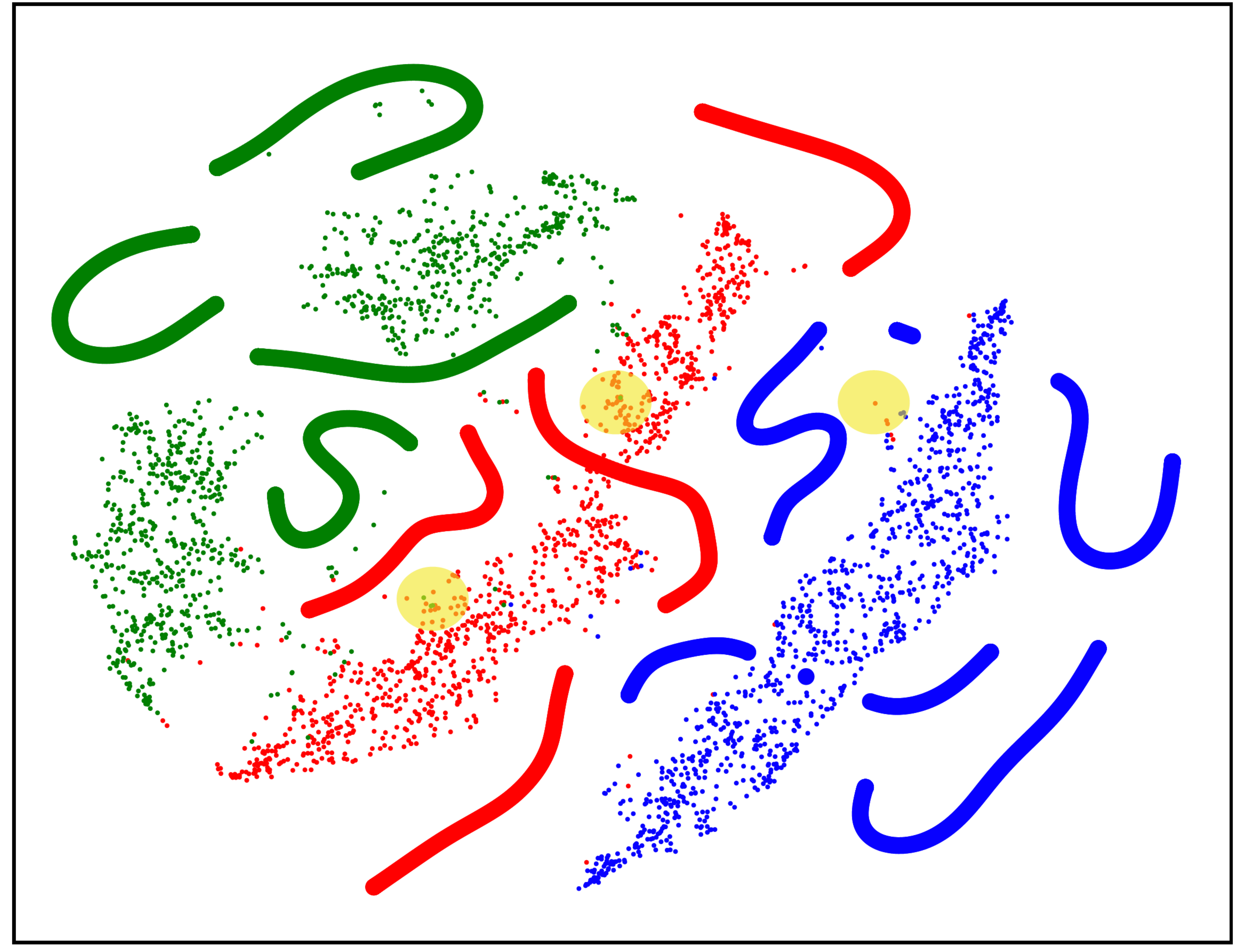}
\caption{Two-dimensional visualizations of the real and generated DE feature (cWGAN) of one subject in the SEED dataset. Data points with red, green and blue colors represent three emotions of negative, neutral and positive, respectively. The lines represent the real data, and thin points represent the generated data. Note that the yellow circles denote bad quality samples.}
\label{fig:7}
\end{figure}

\subsection{Visualization of the Generated Data}

We visualize the generated data with two methods, two-dimensional circular view of the scalp and two-dimensional visualization using t-SNE, to show why our proposed methods work. We selected cWGAN as the generated method and the SEED dataset (DE feature) to represent our results since sVAE and sWGAN have similar vision performance.

Fig. 6 depicts the two-dimensional circular view of the scalp. The generated data have a similar data distribution as the real data. For positive emotion, the lateral areas of both real and generated data are more activated in beta and gamma bands than the other two emotions. For neutral emotion, both the real and generated data had high alpha responses. For negative emotion, high gamma responses at prefrontal sites appeared in real and generated data. These phenomena indicated that our methods can capture the information of the real data distribution. Therefore, the generated samples can be appended to the training set to enhance the performance of the affective models.

As shown in Fig. 7, we plotted the distributions of real and generated DE features (generated by cWGAN) by t-SNE \cite{maaten2008visualizing}. Data from each emotion was clustered in the latent space, and the generated data were close to the corresponding real data, which implies that the generated data carry enough realistic information. This phenomenon also indicates that the data generated by our methods can be used to augment the training dataset.

In addition, the distribution of real data was sparse, and the boundaries of different categories in the data manifold were not obvious. The generated data supplemented the training data manifold, which led to better margins for the classifier. Therefore, we can improve the classification performance by training the classifier with the generated data. We can also explain this phenomenon from another point of view. The generated data have a similar data distribution to the real data, but they are not the real data. Therefore, the generated data not only carry realistic information but also have diverse information. The classifier trained by the augmented data was more robust. This phenomenon is also consistent with the aforementioned classification results.

However, the possibility of generating bad quality samples increased when we added the generating number. This phenomenon occurs no matter what generative methods we apply. For example, we wanted to generate a sample of positive emotions in the SEED dataset, but we might obtain a sample that is more similar to a negative sample by the generative model. We called this sample a bad quality sample. In Fig. 7, we can find some bad quality samples. For example, some generated neutral samples (red points) were more close to the real positive samples (blue lines). In this case, the bad quality sample misled the classifier, and the classification accuracy decreased. We can also find a similar phenomenon in the above tables: the accuracies decayed when too many generated data were appended.

\subsection{Discussions on Different Affective Models}
The abovementioned results show that the performance of the emotion recognition models can be improved by using our proposed data augmentation methods. We achieved performance improvements in different datasets, features, and classifiers, which demonstrates the generalization ability and effectiveness of our methods. Although all three proposed methods improved the performance of EEG-based emotion recognition tasks, they had some differences in terms of stability, accuracy and time usage.

For stability, sVAE had better performance than cWGAN and sWGAN. Although the WGAN had good convergence performance and was more stable than the original GAN, it may collapse because of adversarial training. However, VAE is more stable.

For accuracy, sWGAN had better classification performances than sVAE most of the time. This phenomenon indicates that GAN can capture more latent information than VAE. Therefore, the data generated by GAN are more useful for building the recognition model than those generated by VAE. In addition, sWGAN always performed better than cWGAN on both datasets, which indicates that the selective methods are more efficient at improving emotion recognition models.

For time usage, cWGAN had a quicker convergence speed than sWGAN and sVAE. cWGAN uses all of the generated data without considering their quality, while sWGAN and sVAE need to select the generated data and use the high-quality data to augment the training set. Therefore, the two methods require more computation time to determine the quality of the generated data.

\section{Conclusions}
In this paper, we proposed three deep generative methods for enhancing EEG-based emotion recognition by generating training data. We generated realistic-like PSD and DE features of EEG data with our proposed methods: cWAGN, sVAE, and sWGAN. We augmented the original training dataset using the generated data to improve the accuracy of EEG-based emotion recognition models. The experimental results on two emotion datasets demonstrate the effectiveness of our methods. The emotion recognition models trained on the augmented training datasets achieved 10.2\% and 5.4\% improvements on the SEED dataset and the DEAP dataset, respectively. By visualizing the generated data, we explained the reason for the accuracy improvements. We also studied the performance of the classifiers when adding different numbers of generated data to the original training set. We observed that the classification accuracy decayed when too many generated data were appended. Our experimental results indicate that the number of generated data should be less than 10 times of the original training dataset, and then the affective models achieved the best performance. In addition, we carried out a systematic comparison between the proposed methods. We find that sWGAN had the best performance in terms of accuracy, while it cost more time than cWGAN.

\section{ACKNOWLEDGEMENTS}
This work was supported in part by the National Key Research and Development Program of China (Grant 2017YFB1002501), the National Natural Science Foundation of China (Grant No. 61673266 and No.61976135), the Fundamental Research Funds for the Central Universities, and the 111 Project.

\ifCLASSOPTIONcaptionsoff
  \newpage
\fi

\bibliographystyle{IEEEtran}
\bibliography{ijcai18}

\begin{thebibliography}{10}
\providecommand{\url}[1]{#1}
\csname url@samestyle\endcsname
\providecommand{\newblock}{\relax}
\providecommand{\bibinfo}[2]{#2}
\providecommand{\BIBentrySTDinterwordspacing}{\spaceskip=0pt\relax}
\providecommand{\BIBentryALTinterwordstretchfactor}{4}
\providecommand{\BIBentryALTinterwordspacing}{\spaceskip=\fontdimen2\font plus
\BIBentryALTinterwordstretchfactor\fontdimen3\font minus
  \fontdimen4\font\relax}
\providecommand{\BIBforeignlanguage}[2]{{%
\expandafter\ifx\csname l@#1\endcsname\relax
\typeout{** WARNING: IEEEtran.bst: No hyphenation pattern has been}%
\typeout{** loaded for the language `#1'. Using the pattern for}%
\typeout{** the default language instead.}%
\else
\language=\csname l@#1\endcsname
\fi
#2}}
\providecommand{\BIBdecl}{\relax}
\BIBdecl

\bibitem{emotionAI1}
M.~Somers, ``Emotion {AI}, explained,''
  \url{https://mitsloan.mit.edu/ideas-made-to-matter/emotion-ai-explained },
  2019.

\bibitem{emotionAI2}
D.~Smith and B.~Burke, ``Hype cycle for emerging technologies, 2019,''
  \url{https://www.gartner.com/en/documents/3956015/hype-cycle-for-emerging-technologies-2019
  }, 2019.

\bibitem{Bocharov2017Depression}
A.~V. Bocharov, G.~G. Knyazev, and A.~N. Savostyanov, ``Depression and implicit
  emotion processing: An {EEG} study,'' \emph{Neurophysiologie Clinique},
  vol.~47, no.~3, p. S0987705316303513, 2017.

\bibitem{Liu2017A}
Z.~Liu, W.~Min, W.~Cao, L.~Chen, J.~Xu, R.~Zhang, M.~Zhou, and J.~Mao, ``A
  facial expression emotion recognition based human-robot interaction system,''
  \emph{IEEE/CAA Journal of Automatica Sinica}, vol.~4, no.~4, pp. 668--676,
  2017.

\bibitem{Garber2012Using}
M.~Garber-Barron and M.~Si, ``Using body movement and posture for emotion
  detection in non-acted scenarios,'' in \emph{IEEE International Conference on
  Fuzzy Systems}, 2012, pp. 1--8.

\bibitem{Tanja2009Emotion}
B.~Tanja, G.~Didier, and K.~R. Scherer, ``Emotion recognition from expressions
  in face, voice, and body: the multimodal emotion recognition test (mert),''
  \emph{Emotion}, vol.~9, no.~5, p. 691, 2009.

\bibitem{Samara2017Feature}
A.~Samara, M.~L.~R. Menezes, and L.~Galway, ``Feature extraction for emotion
  recognition and modelling using neurophysiological data,'' in
  \emph{International Conference on Ubiquitous Computing \& Communications \&
  International Symposium on Cyberspace \& Security}, 2017, pp. 138--144.

\bibitem{Wang2014Emotional}
X.-W. Wang, D.~Nie, and B.-L. Lu, ``Emotional state classification from eeg
  data using machine learning approach,'' \emph{Neurocomputing}, vol. 129, no.
  apr.10, pp. 94--106, 2014.

\bibitem{McFarland_2016}
D.~J. McFarland, M.~A. Parvaz, W.~A. Sarnacki, R.~Z. Goldstein, and J.~R.
  Wolpaw, ``Prediction of subjective ratings of emotional pictures by {EEG}
  features,'' \emph{Journal of Neural Engineering}, vol.~14, no.~1, p. 016009,
  dec 2016.

\bibitem{Alarc1949Emotions}
S.~Meneses~Alarcão and M.~J. Fonseca, ``Emotions recognition using {EEG}
  signals: A survey,'' \emph{IEEE Transactions on Affective Computing},
  vol.~10, no.~3, pp. 374--393, 2017.

\bibitem{Craik_2019}
A.~Craik, Y.~He, and J.~L. Contreras-Vidal, ``Deep learning for
  electroencephalogram ({EEG}) classification tasks: a review,'' \emph{Journal
  of Neural Engineering}, vol.~16, no.~3, p. 031001, apr 2019.

\bibitem{9043472}
R.~{Fourati}, B.~{Ammar}, J.~{Sanchez-Medina}, and A.~M. {Alimi},
  ``Unsupervised learning in reservoir computing for eeg-based emotion
  recognition,'' \emph{IEEE Transactions on Affective Computing}, pp. 1--1,
  2020.

\bibitem{zheng2015investigating}
W.-L. Zheng and B.-L. Lu, ``Investigating critical frequency bands and channels
  for {EEG}-based emotion recognition with deep neural networks,'' \emph{IEEE
  Transactions on Autonomous Mental Development}, vol.~7, no.~3, pp. 162--175,
  2015.

\bibitem{koelstra2012deap}
S.~Koelstra, C.~M{\"u}hl, M.~Soleymani, J.-S. Lee, A.~Yazdani, T.~Ebrahimi,
  T.~Pun, A.~Nijholt, and I.~Patras, ``{DEAP}: A database for emotion analysis;
  using physiological signals,'' \emph{IEEE Transactions on Affective
  Computing}, vol.~3, no.~1, pp. 18--31, 2012.

\bibitem{7887697}
S.~{Katsigiannis} and N.~{Ramzan}, ``Dreamer: A database for emotion
  recognition through eeg and ecg signals from wireless low-cost off-the-shelf
  devices,'' \emph{IEEE Journal of Biomedical and Health Informatics}, vol.~22,
  no.~1, pp. 98--107, Jan 2018.

\bibitem{5975141}
M.~{Soleymani}, J.~{Lichtenauer}, T.~{Pun}, and M.~{Pantic}, ``A multimodal
  database for affect recognition and implicit tagging,'' \emph{IEEE
  Transactions on Affective Computing}, vol.~3, no.~1, pp. 42--55, Jan 2012.

\bibitem{8606087}
T.~{Song}, W.~{Zheng}, C.~{Lu}, Y.~{Zong}, X.~{Zhang}, and Z.~{Cui}, ``Mped: A
  multi-modal physiological emotion database for discrete emotion
  recognition,'' \emph{IEEE Access}, vol.~7, pp. 12\,177--12\,191, 2019.

\bibitem{UnreasonableData}
A.~{Halevy}, P.~{Norvig}, and F.~{Pereira}, ``The unreasonable effectiveness of
  data,'' \emph{IEEE Intelligent Systems}, vol.~24, no.~2, pp. 8--12, March
  2009.

\bibitem{Zhang2018A}
Q.~Zhang, L.~T. Yang, Z.~Chen, and L.~Peng, ``A survey on deep learning for big
  data,'' \emph{Information Fusion}, vol.~42, pp. 146--157, 2018.

\bibitem{Krell2017Rotational}
M.~M. {Krell} and S.~K. {Kim}, ``Rotational data augmentation for
  electroencephalographic data,'' in \emph{2017 39th Annual International
  Conference of the IEEE Engineering in Medicine and Biology Society (EMBC)},
  July 2017, pp. 471--474.

\bibitem{Lotte2015Signal}
F.~Lotte, ``Signal processing approaches to minimize or suppress calibration
  time in oscillatory activity-based brain–-computer interfaces,''
  \emph{Proceedings of the IEEE}, vol. 103, no.~6, pp. 871--890, 2015.

\bibitem{Fang2018Data}
F.~Wang, S.-H. Zhong, J.-F. Peng, J.-M. Jiang, and Y.~Liu, ``Data augmentation
  for {EEG}-based emotion recognition with deep convolutional neural
  networks,'' in \emph{International Conference on Multimedia Modeling}, 2018,
  pp. 82--93.

\bibitem{Hartmann2018EEG}
K.~G. Hartmann, R.~T. Schirrmeister, and T.~Ball, ``{EEG-GAN}: Generative
  adversarial networks for electroencephalograhic ({EEG}) brain signals,''
  \emph{arXiv preprint arXiv:1806.01875}, 2018.

\bibitem{luo2018embc}
Y.~Luo and B.-L. Lu, ``{EEG} data augmentation for emotion recognition using a
  conditional {Wasserstein} {GAN},'' in \emph{2018 40th Annual International
  Conference of the IEEE Engineering in Medicine and Biology Society
  (EMBC)}.\hskip 1em plus 0.5em minus 0.4em\relax IEEE, 2018, pp. 2535--2538.

\bibitem{duan2013differential}
R.-N. Duan, J.-Y. Zhu, and B.-L. Lu, ``Differential entropy feature for
  {EEG}-based emotion classification,'' in \emph{International IEEE/EMBS
  Conference on Neural Engineering (NER)}.\hskip 1em plus 0.5em minus
  0.4em\relax IEEE, 2013, pp. 81--84.

\bibitem{yang2018eeg-based}
Y.~Yang, Q.~M.~J. Wu, W.-l. Zheng, and B.-l. Lu, ``{EEG}-based emotion
  recognition using hierarchical network with subnetwork nodes,'' \emph{IEEE
  Transactions on Cognitive and Developmental Systems}, vol.~10, no.~2, pp.
  408--419, 2018.

\bibitem{kingma2014auto-encoding}
D.~P. Kingma and M.~Welling, ``Auto-encoding variational bayes,'' in
  \emph{International Conference on Learning Representations}, 2014.

\bibitem{arjovsky2017wasserstein}
M.~Arjovsky, S.~Chintala, and L.~Bottou, ``Wasserstein {GAN},'' \emph{arXiv
  preprint arXiv:1701.07875}, 2017.

\bibitem{gulrajani2017improved}
I.~Gulrajani, F.~Ahmed, M.~Arjovsky, V.~Dumoulin, and A.~C. Courville,
  ``Improved training of {Wasserstein} {GANs},'' in \emph{Neural Information
  Processing Systems}, 2017, pp. 5769--5779.

\bibitem{Zander2012Context}
T.~O. Zander and S.~Jatzev, ``Context-aware brain-computer interfaces:
  exploring the information space of user, technical system and environment.''
  \emph{Journal of Neural Engineering}, vol.~9, no.~1, p. 016003, 2011.

\bibitem{M2014A}
C.~M{\"u}hl, B.~Allison, A.~Nijholt, and G.~Chanel, ``A survey of affective
  brain computer interfaces: principles, state-of-the-art, and challenges,''
  \emph{Brain-Computer Interfaces}, vol.~1, no.~2, pp. 66--84, 2014.

\bibitem{alarcao2017emotions}
S.~M. Alarcao and M.~J. Fonseca, ``Emotions recognition using {EEG} signals: A
  survey,'' \emph{IEEE Transactions on Affective Computing}, vol.~10, no.~3,
  pp. 374--393, 2017.

\bibitem{jenke2014feature}
R.~Jenke, A.~Peer, and M.~Buss, ``Feature extraction and selection for emotion
  recognition from {EEG},'' \emph{IEEE Transactions on Affective Computing},
  vol.~5, no.~3, pp. 327--339, 2014.

\bibitem{highOrder}
P.~C. {Petrantonakis} and L.~J. {Hadjileontiadis}, ``Emotion recognition from
  {EEG} using higher order crossings,'' \emph{IEEE Transactions on Information
  Technology in Biomedicine}, vol.~14, no.~2, pp. 186--197, March 2010.

\bibitem{music1}
Y.~{Lin}, C.~{Wang}, T.~{Jung}, T.~{Wu}, S.~{Jeng}, J.~{Duann}, and J.~{Chen},
  ``{EEG}-based emotion recognition in music listening,'' \emph{IEEE
  Transactions on Biomedical Engineering}, vol.~57, no.~7, pp. 1798--1806, July
  2010.

\bibitem{music2}
O.~Sourina, Y.~Liu, and M.~K. Nguyen, ``Real-time {EEG}-based emotion
  recognition for music therapy,'' \emph{Journal on Multimodal User
  Interfaces}, vol.~5, pp. 27--35, 2012.

\bibitem{sohaib2013evaluating}
A.~T. Sohaib, S.~Qureshi, J.~Hagelback, O.~Hilborn, and P.~Jercic, ``Evaluating
  classifiers for emotion recognition using {EEG},'' in \emph{International
  Conference on Augmented Cognition}, 2013, pp. 492--501.

\bibitem{petrantonakis2011a}
P.~C. Petrantonakis and L.~J. Hadjileontiadis, ``A novel emotion elicitation
  index using frontal brain asymmetry for enhanced {EEG}-based emotion
  recognition,'' \emph{Engineering in Medicine and Biology Society}, vol.~15,
  no.~5, pp. 737--746, 2011.

\bibitem{zheng2017identifying}
W.-L. Zheng, J.-Y. Zhu, and B.-L. Lu, ``Identifying stable patterns over time
  for emotion recognition from {EEG},'' \emph{IEEE Transactions on Affective
  Computing}, pp. 1--1, 2017.

\bibitem{zheng2019emotionmeter}
W.-L. Zheng, W.~Liu, Y.-F. Lu, B.-L. Lu, and A.~Cichocki, ``Emotionmeter: A
  multimodal framework for recognizing human emotions,'' \emph{IEEE
  Transactions on Cybernetics}, vol.~49, no.~3, pp. 1110--1122, March 2019.

\bibitem{zhao2018emotion}
L.-M. Zhao, R.~Li, W.-L. Zheng, and B.-L. Lu, ``Classification of five emotions
  from {EEG} and eye movement signals: complementary representation
  properties,'' in \emph{International IEEE/EMBS Conference on Neural
  Engineering (NER)}.\hskip 1em plus 0.5em minus 0.4em\relax IEEE, 2019, pp.
  611--614.

\bibitem{goodfellow2014generative}
I.~Goodfellow, J.~Pouget-Abadie, M.~Mirza, B.~Xu, D.~Warde-Farley, S.~Ozair,
  A.~Courville, and Y.~Bengio, ``Generative adversarial nets,'' in \emph{Neural
  Information Processing Systems}, 2014, pp. 2672--2680.

\bibitem{salimans2015markov}
T.~Salimans, D.~Kingma, and M.~Welling, ``Markov chain monte carlo and
  variational inference: Bridging the gap,'' in \emph{International Conference
  on Machine Learning}, 2015, pp. 1218--1226.

\bibitem{kulkarni2015deep}
T.~D. Kulkarni, W.~F. Whitney, P.~Kohli, and J.~Tenenbaum, ``Deep convolutional
  inverse graphics network,'' in \emph{Neural Information Processing Systems},
  2015, pp. 2539--2547.

\bibitem{gregor2015draw}
K.~Gregor, I.~Danihelka, A.~Graves, D.~J. Rezende, and D.~Wierstra, ``Draw: A
  recurrent neural network for image generation,'' \emph{arXiv preprint
  arXiv:1502.04623}, 2015.

\bibitem{Mirza2014Conditional}
M.~Mirza and S.~Osindero, ``Conditional generative adversarial nets,''
  \emph{arXiv preprint arXiv:1411.1784}, 2014.

\bibitem{infoGAN}
X.~Chen, Y.~Duan, R.~Houthooft, J.~Schulman, I.~Sutskever, and P.~Abbeel,
  ``Info{GAN}: Interpretable representation learning by information maximizing
  generative adversarial nets,'' \emph{arXiv preprint arXiv:1606.03657v1},
  2016.

\bibitem{ledig2017photo-realistic}
C.~Ledig, L.~Theis, F.~Huszar, J.~Caballero, A.~Cunningham, A.~Acosta, A.~P.
  Aitken, A.~Tejani, J.~Totz, Z.~Wang \emph{et~al.}, ``Photo-{Realistic} single
  image super-resolution using a generative adversarial network,'' in
  \emph{Computer {Vision} and {Pattern} {Recognition}}, 2017, pp. 105--114.

\bibitem{wu2016learning}
J.~Wu, C.~Zhang, T.~Xue, W.~T. Freeman, and J.~B. Tenenbaum, ``Learning a
  probabilistic latent space of object shapes via 3d generative-adversarial
  modeling,'' in \emph{Neural Information Processing Systems}, 2016, pp.
  82--90.

\bibitem{Li2017Adversarial}
J.~Li, W.~Monroe, T.~Shi, S.~Jean, A.~Ritter, and J.~Dan, ``Adversarial
  learning for neural dialogue generation,'' \emph{arXiv preprint
  arXiv:1701.06547}, 2017.

\bibitem{Choi2017Generating}
E.~Choi, S.~Biswal, B.~Malin, J.~Duke, W.~F. Stewart, and J.~Sun, ``Generating
  multi-label discrete electronic health records using generative adversarial
  networks,'' \emph{arXiv preprint arXiv:1703.06490}, 2017.

\bibitem{Mogren2016C}
O.~Mogren, ``C-{RNN}-{GAN}: Continuous recurrent neural networks with
  adversarial training,'' \emph{arXiv preprint arXiv:1611.09904}, 2016.

\bibitem{Radford2015Unsupervised}
A.~Radford, L.~Metz, and S.~Chintala, ``Unsupervised representation learning
  with deep convolutional generative adversarial networks,'' in
  \emph{International Conference on Learning Representations}, 2016.

\bibitem{began}
D.~Berthelot, T.~Schumm, and L.~Metz, ``{BEGAN}: Boundary equilibrium
  generative adversarial networks,'' \emph{arXiv preprint arXiv:1703.10717},
  2017.

\bibitem{Qi2017Loss}
G.~J. Qi, ``Loss-sensitive generative adversarial networks on lipschitz
  densities,'' \emph{arXiv preprint arXiv:1701.06264v6}, 2017.

\bibitem{artData}
D.~A. van Dyk and X.-L. Meng, ``The art of data augmentation,'' \emph{Journal
  of Computational and Graphical Statistics}, vol.~10, no.~1, pp. 1--50, 2001.

\bibitem{Krizhevsky2012ImageNet}
A.~Krizhevsky, I.~Sutskever, and G.~E. Hinton, ``Imagenet classification with
  deep convolutional neural networks,'' in \emph{Advances in Neural Information
  Processing Systems}, 2012, pp. 1097--1105.

\bibitem{Perez2017The}
L.~Perez and J.~Wang, ``The effectiveness of data augmentation in image
  classification using deep learning,'' \emph{arXiv preprint arXiv:1712.04621},
  2017.

\bibitem{simard2003best}
P.~Y. Simard, D.~W. Steinkraus, and J.~Platt, ``Best practices for
  convolutional neural networks applied to visual document analysis,'' in
  \emph{International Conference on Document Analysis and Recognition}, 2003,
  pp. 958--963.

\bibitem{zheng2017unlabeled}
Z.~Zheng, L.~Zheng, and Y.~Yang, ``Unlabeled samples generated by gan improve
  the person re-identification baseline in vitro,'' in \emph{Proceedings of the
  IEEE International Conference on Computer Vision}, 2017, pp. 3754--3762.

\bibitem{zhu2018emotion}
X.~Zhu, Y.~Liu, J.~Li, T.~Wan, and Z.~Qin, ``Emotion classification with data
  augmentation using generative adversarial networks,'' in \emph{Pacific-Asia
  Conference on Knowledge Discovery and Data Mining}.\hskip 1em plus 0.5em
  minus 0.4em\relax Springer, 2018, pp. 349--360.

\bibitem{hinton2006reducing}
G.~E. Hinton and R.~R. Salakhutdinov, ``Reducing the dimensionality of data
  with neural networks,'' \emph{Science}, vol. 313, no. 5786, pp. 504--507,
  2006.

\bibitem{simonyan2014very}
K.~Simonyan and A.~Zisserman, ``Very deep convolutional networks for
  large-scale image recognition,'' \emph{arXiv preprint arXiv:1409.1556}, 2014.

\bibitem{mikolov2013efficient}
T.~Mikolov, K.~Chen, G.~Corrado, and J.~Dean, ``Efficient estimation of word
  representations in vector space,'' \emph{arXiv preprint arXiv:1301.3781},
  2013.

\bibitem{dahl2011context}
G.~E. Dahl, D.~Yu, L.~Deng, and A.~Acero, ``Context-dependent pre-trained deep
  neural networks for large-vocabulary speech recognition,'' \emph{IEEE
  Transactions on Audio, Speech and Language Processing}, vol.~20, no.~1, pp.
  30--42, 2011.

\bibitem{he2016deep}
K.~He, X.~Zhang, S.~Ren, and J.~Sun, ``Deep residual learning for image
  recognition,'' in \emph{Proceedings of the IEEE Conference on Computer Vision
  and Pattern Recognition}, 2016, pp. 770--778.

\bibitem{Shi2013Differential}
L.~C. Shi, Y.~Y. Jiao, and B.~L. Lu, ``Differential entropy feature for
  {EEG}-based vigilance estimation,'' \emph{International IEEE/EMBS Conference
  on Neural Engineering (NER)}, vol. 2013, no. 2013, pp. 6627--6630, 2013.

\bibitem{maaten2008visualizing}
L.~V.~D. Maaten and G.~Hinton, ``Visualizing data using {t-SNE},''
  \emph{Journal of Machine Learning Research}, vol.~9, no. Nov, pp. 2579--2605,
  2008.

\end{thebibliography}

\end{document}